\documentclass[twocolumn,aps,nofootinbib,longbibliography]{revtex4-2}
\usepackage[dvipdfmx]{graphicx}
\usepackage{mathtools,amssymb,amsmath,bm,url,color}
\usepackage[breaklinks]{hyperref}
\usepackage{float}
\usepackage{multirow}
\allowdisplaybreaks[1]

\definecolor{myred}{RGB}{200,50,50}

\begin{document}

\title{Spin excitation spectra in helimagnetic states: proper-screw, cycloid, vortex crystal, and hedgehog lattice}

\author{Yasuyuki~Kato, Satoru~Hayami,  and Yukitoshi~Motome}

\affiliation{Department of Applied Physics, the University of Tokyo, Tokyo 113-8656, Japan}

\begin{abstract}
\noindent
We investigate the spin excitation spectra in chiral and polar magnets by the linear spin-wave theory for an effective spin model with symmetric and antisymmetric long-range interactions.
In one dimension, we obtain the analytic form of the dynamical spin structure factor for proper-screw and cycloidal helical spin states with uniform twists, 
which shows a gapless mode with strong intensity at the helical wave number.
When introducing spin anisotropy in the symmetric interactions, we 
numerically show that the stable spin spirals become elliptically anisotropic with nonuniform twists and the spin excitation is gapped.
In higher dimensions, we find that similar anisotropy stabilizes multiple-$Q$ spin states, such as vortex crystals and hedgehog lattices. 
We show that the anisotropy in these states manifests itself in the dynamical spin structure factor:
a strong intensity in the transverse components to the wave number 
appears only when the helical wave vector and the corresponding easy axis are perpendicular to each other.
Our findings could be useful not only to identify the spin structure but also to deduce the stabilization mechanism by inelastic neutron scattering measurements.
\end{abstract}


\maketitle

\section{Introduction}

\begin{figure}[bh]
  \centering
  \includegraphics[trim=0 0 0 0, clip,width=\columnwidth]{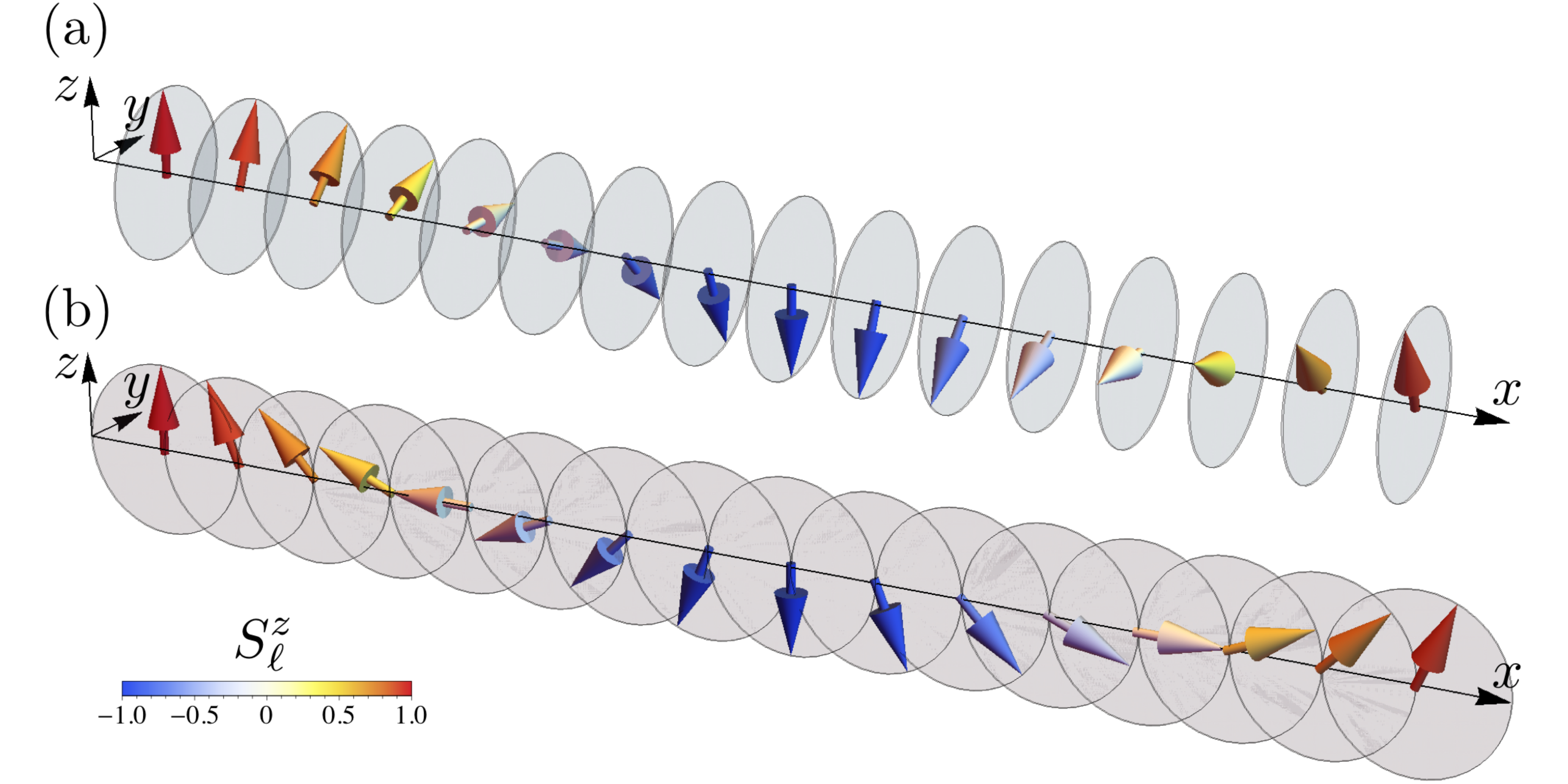}
  \caption{Helimagnetic orders: (a) proper-screw and (b) cycloidal helical spin states. The latter is obtained by $\pi/2$ spin rotation of the former about the $z$ axis.}
  \label{fig01}
\end{figure}

The helimagnetic orders are periodic spin states found in a wide range of materials, from metals to insulators, where the magnetic moments form 
twisting and swirling textures, such as spin spirals and vortex crystals (VCs)~\cite{Bogdanov1989}. 
Of particular interest is the cases where the spin textures define topologically nontrivial objects~\cite{Braun2012,Seidel2016,Bogdanov2020,Tokura2020}.
There are many examples of such helimagnetic orders, 
e.g., one-dimensional (1D) proper-screw helical spin (HS) states~[Fig.~\ref{fig01}(a)]~\cite{Yoshimori1959}, 
1D cycloidal HS states~[Fig.~\ref{fig01}(b)]~\cite{Moriya1960}, 
1D chiral soliton lattice~\cite{Dzyaloshinskii1964,Dzyaloshinskii1965,Dzyaloshinskii1965b,Izyumov1984,Kishine2009,Togawa2012,Togawa2016}, 
two-dimensional (2D) skyrmion crystals (SkXs)~\cite{Muhlbauer2009,Yu2010,Nagaosa2013,Fert2017}, 
2D vortex crystal (VC)~\cite{Khanh2020},
and three-dimensional (3D) hedgehog lattices (HLs)~\cite{Tanigaki2015,Kanazawa2016,Kanazawa2017,Fujishiro2019,Fujishiro2020,Kanazawa2020}. 
These helimagnetic states have been attracting a lot of attention since they induce intriguing electronic and transport properties, such as
the magnetoelectric effect~\cite{Tokura2010} and the topological Hall effect~\cite{Nagaosa2010}, which would lay the cornerstone of future technology.

Several mechanisms have been proposed for the stability of these helimagnetic spin textures, including the Dzyaloshinskii-Moriya antisymmetric exchange interactions~\cite{Rossler2006,Yi2009,Okumura2017}, 
frustration among the competing exchange interactions~\cite{Okubo2012,Leonov2015,Lin2016}, 
four spin interactions~\cite{Momoi1997,Kurz2001,Heinze2011,Brinker2019,Laszloffy2019,Paul2020}, 
long-range interactions via itinerant electrons~\cite{Martin2008,Akagi2010,Kato2010,Akagi2012,Ozawa2017,Hayami2017,Hayami2019,Wang2020,Hayami2021b}, 
long-range dipole interactions~\cite{Ezawa2010,Kwon2012,Utesov2021}, 
and bond-dependent anisotropic interaction~\cite{Hayami2020,Hayami2021,Wang2021}.
To elucidate the relevant mechanism, 
it is desired to clarify the microscopic information of the magnetic interactions. 
Inelastic neutron scattering is a useful experimental tool to obtain such microscopic information from the analysis of the spin excitation spectrum.
It is, however, not always an easy task, especially for the complex spin textures. 
For example, while the SkXs and the HLs are stably obtained for the models with either short-range~\cite{Bak1980,Rossler2006,Okubo2012,Yang2016} 
or long-range interactions~\cite{Ozawa2017,Hayami2017,Hayami2018,Hayami2019,Okumura2020}, 
it remains yet to be clarified which is the most relevant mechanism in each substance. 
This is mainly due to less available information on the spin excitations for the detailed comparison between theory and experiment.

In this paper, we systematically study the spin excitation spectra for spin models which stabilize various types of helimagnetic spin textures, by tuning the range of magnetic interactions in real space. 
Specifically, starting from the effective spin model for spin-charge coupled systems, which has infinite-range interactions~\cite{Hayami2017,Hayami2018}, we extend it by including both symmetric and antisymmetric exchange interactions with spatial decay, and obtain the ground states and spin excitation spectra by 
variational calculations and the linear spin-wave theory, respectively. 
We find that our models stabilize 2D VCs and 3D HLs in addition to 1D HS states, by introducing spin anisotropy in the symmetric interaction.
In the 1D case, we show that the dynamical spin structure factor for both proper-screw and cycloidal HS states has a gapless mode with strong intensity at the helical wave number in the isotropic case, but they are gapped in the presence of the anisotropy which modulates the stable spin spirals into elliptically anisotropic ones and makes the twists nonuniform. 
We also clarify that the lowest-energy excitation mode with the strongest intensity can be regarded as a phase shift of the spin helix.
In higher dimensions, we find that while the system exhibits a HS state in the isotropic case, the anisotropy can stabilize
multiple-$Q$ spin states which are composed of superpositions of multiple spin helices; 
we obtain four different types of double-$Q$ ($2Q$) VCs in two dimensions and three different types of triple-$Q$ ($3Q$) HLs in three dimensions.
We find that the dynamical spin structure factor for the multiple-$Q$ spin states exhibits a strong intensity in the lowest-energy excitation mode 
when the helical wave vector is perpendicular to the easy axis of the corresponding interaction. 
This means that the experimental identification of such strong intensity by the inelastic neutron scattering would provide the information of not only the propagating direction and magnetic period of the helices
but also the anisotropy in the effective magnetic interactions.
In addition to the experimental relevance, 
our present scheme provides a versatile theoretical framework to investigate spin-wave excitations in a wide variety of multiple-$Q$ spin states, even beyond those treated in this paper, such as SkXs and other multiple-$Q$ HLs.

The structure of this paper is as follows.
In Sec.~\ref{sec:model}, we first introduce the effective spin model for chiral magnets with infinite-range interactions.
Several types of the symmetric and antisymmetric interactions are introduced for the 1D, 2D, and 3D cases.
Then, we extend the model by introducing spatial decay in the interactions.
In Sec.~\ref{sec:method}, we describe the methods used in the present study:
the variational method for the ground state and the linear spin-wave theory for the spin excitations.
In Sec.~\ref{sec:results}, the results for the 1D, 2D, and 3D cases are shown.
For the 1D case, we present the 
results of the analytical calculations for the HS states with spatially uniform spin twist in the isotropic case, and the results of the numerical calculations for the effect of the anisotropy.
For the 2D and 3D cases, we show the ground-state phase diagrams while changing the anisotropy in the symmetric interaction and the strength of the antisymmetric interaction.
Then, we discuss the details of the stabilized spin states, the spin-wave dispersion, and the dynamical spin structure factor, which is relevant to the inelastic neutron scattering experiments, for different types of VCs and HLs.
Section~\ref{sec:summary_discussions} is devoted to the summary and discussion.

\section{Model}\label{sec:model}

\subsection{Effective spin model}

We begin with a generic spin model for chiral magnets whose
Hamiltonian is defined in momentum space as
\begin{align}
\mathcal{H}=& \sum_{{\bf q}\in 1{\rm BZ}}\mathcal{H}_{{\bf q}},
\label{eq:H}
\end{align}
with
\begin{align}
\mathcal{H}_{{\bf q}} =&
- \sum_{\alpha,\beta}J^{\alpha\beta}_{{\bf q}}
{S}^\alpha_{{\bf q}} 
{S}^\beta_{-{\bf q}}
- i {\bf D}_{{\bf q}}
\cdot
\left(
{\bf S}_{{\bf q}} \times
{\bf S}_{-{\bf q}} \right),
\label{eq:Hq}
\end{align}
where $\alpha$, $\beta = x$, $y$, $z$; 
${\bf S}_{\bf q} = (S^x_{\bf q}, S^y_{\bf q}, S^z_{\bf q})$ is defined by the Fourier transform
of the spin in real space, ${\bf S}_{\bf r}$, as
\begin{align}
{\bf S}_{\bf q} = L^{-\frac{d}{2}} \sum_{\bf r} {\bf S}_{\bf r} e^{- i {\bf q}\cdot{\bf r}}.
\end{align}
Here we define this model on a $d$-dimensional hypercubic lattice with linear dimension $L$ under the periodic boundary condition; 
the lattice site ${\bf r}$ is denoted as 
\begin{align}
{\bf r}=
\begin{cases}
 x \;\; (\equiv \ell) , & (d=1)\\
 (x ,y) , & (d=2)\\
 (x , y, z)  , & (d=3)
\end{cases}
,
\end{align}
with integers $x$, $y$, and $z$ in $[0,L)$.
The sum $\sum_{{\bf q}\in{\rm 1BZ}}$ in Eq.~\eqref{eq:H} runs over all the wave numbers in the first Brillouin zone (${\rm 1BZ}$):
\begin{align}
{\bf q}=
\begin{cases}
q_x \equiv q= \frac{2\pi}{L} n_x, & (d=1)\\
(q_x,q_y)= \frac{2\pi}{L} (n_x, n_y)        ,& (d=2)\\
(q_x,q_y,q_z)= \frac{2\pi}{L} (n_x, n_y, n_z) ,& (d=3)
\end{cases},
\end{align}
with integers $n_\alpha$ in $[-L/2,L/2)$, that is, $-\pi \leq q_{\alpha} < \pi$.
The first term of $\mathcal{H}_{\bf q}$ in Eq.~\eqref{eq:Hq} represents the symmetric exchange interaction ($J_{\bf q}^{\alpha\beta}=J_{\bf q}^{\beta\alpha}$), 
while the second term represents the antisymmetric one of the Dzyaloshinskii-Moriya type~\cite{Dzyaloshinsky1958,Moriya1960}. 
For the former,
we include only the diagonal elements, namely, $J_{\bf q}^{\alpha \beta} = J_{\bf q}^{\alpha\alpha} \delta_{\alpha , \beta}$, for simplicity ($\delta_{\alpha,\beta}$ is the Kronecker delta).
Then, $\mathcal{H}_{\bf q}$ is expressed as
\begin{align}
\mathcal{H}_{\bf q} = - \sum_{\alpha,\beta}
S^{\alpha}_{\bf q} \mathcal{J}_{\bf q}^{\alpha \beta} S^{\beta}_{-{\bf q}} ,
\end{align}
with
\begin{align}
\mathcal{J}_{\bf q} = 
\begin{bmatrix}
J_{\bf q}^{xx}  &  i D_{\bf q}^z   & - i D_{\bf q}^y\\
-i D_{\bf q}^z  &  J_{\bf q}^{yy}  & i D_{\bf q}^x \\
 i D_{\bf q}^y  & -i D_{\bf q}^x   &  J_{\bf q}^{zz} 
\end{bmatrix}
=\mathcal{J}^*_{-{\bf q}}.
\label{eq:J_q_matrix}
\end{align} 

\subsection{Infinite-range limit}
\label{subsec:Infinite-range limit}
\begin{table*}
\caption{\label{tab1}
Theoretical models in the present study: dimension $d$, target spin states, spin configurations, equations and schematics of the symmetric and antisymmetric interactions, 
crystallographic point groups, and 
corresponding sections for the results.
}
\begin{ruledtabular}
\begin{tabular}{llllllll}
$d$ & target spin state & spin configuration & \centering{$J_{{\bf Q}_\eta}^{\alpha\alpha}$} & ${\bf D}_{{\bf Q}_\eta}$ & schematic & crystallographic point group & results \\
\hline
\multirow{3}{*}{1D}& 
proper-screw HS state &Figs.~\ref{fig11}(a)--\ref{fig11}(c) & Eq.~\eqref{eq:anisotropy} & Eq.~\eqref{eq:Dq1d} & Fig.~\ref{fig02}(a)  & orthorhombic $D_2$ ($222$) & \multirow{3}{*}{Sec.~\ref{sec:result:1d}}\\
&cycloid(I) HS state & & Eq.~\eqref{eq:anisotropy2} & Eq.~\eqref{eq:Dq1d2} & Fig.~\ref{fig02}(b) & orthorhombic  $C_{2v}$ ($mm2$)  &\\
&cycloid(II) HS state & & Eq.~\eqref{eq:anisotropy3} & Eq.~\eqref{eq:Dq1d2} & Fig.~\ref{fig02}(c) & orthorhombic  $C_{2v}$ ($mm2$) & \vspace{0.2cm}\\
\multirow{4}{*}{2D}&
proper-screw(I) VC & Fig.~\ref{fig15}(a) & Eq.~\eqref{eq:JQeta2d1} & Eq.~\eqref{eq:DQeta2d} & Fig.~\ref{fig03}(a) & tetragonal $D_4$ ($422$)  & \multirow{4}{*}{Sec.~\ref{sec:result2d}}\\
&cycloid(I) VC  & Fig.~\ref{fig15}(b) & Eq.~\eqref{eq:JQeta2d2} & Eq.~\eqref{eq:DQeta2d2} & Fig.~\ref{fig03}(b) & tetragonal $C_{4v}$ ($4mm$) & \\
&proper-screw(II) VC & Fig.~\ref{fig15}(c) & Eq.~\eqref{eq:JQeta2d3} & Eq.~\eqref{eq:DQeta2d3} & Fig.~\ref{fig03}(c) & tetragonal $D_{2d}$ ($\bar{4}2m$) & \\
&cycloid(II) VC  & Fig.~\ref{fig15}(d) & Eq.~\eqref{eq:JQeta2d4} & Eq.~\eqref{eq:DQeta2d4} & Fig.~\ref{fig03}(d) & tetragonal $D_{2d}$ ($\bar{4}m2$) & \vspace{0.2cm}\\
\multirow{3}{*}{3D}&
proper-screw HL & Fig.~\ref{fig19}(a) & Eq.~\eqref{eq:JQeta3d1} & Eq.~\eqref{eq:DQeta3d} & Fig.~\ref{fig04}(a) & cubic $T$ ($23$) & \multirow{3}{*}{Sec.~\ref{sec:result3d}}\\
&cycloid(I) HL & Fig.~\ref{fig19}(b) & Eq.~\eqref{eq:JQeta3d2}  & Eq.~\eqref{eq:DQeta3d2} & Fig.~\ref{fig04}(b) & trigonal $C_3$ ($3$) & \\
&cycloid(II) HL  & Fig.~\ref{fig19}(c) &  Eq.~\eqref{eq:JQeta3d3}  & Eq.~\eqref{eq:DQeta3d3} & Fig.~\ref{fig04}(c) &  trigonal $C_3$ ($3$) & \\
\end{tabular}
\end{ruledtabular}
\end{table*}
A particular case of the model in Eq.~\eqref{eq:H} was studied for 2D VCs and SkXs~\cite{Hayami2018}, where $J^{\alpha \alpha}_{\bf q}$ and ${\bf D}_{\bf q}$ in Eq.~\eqref{eq:J_q_matrix} are taken as
\begin{align}
\label{eq:J_q_infinite}
&J^{\alpha \alpha}_{\bf q} = 
\sum_\eta J^{\alpha \alpha}_{{\bf Q}_\eta} 
(\delta_{{\bf q} , {\bf Q}_\eta}+\delta_{{\bf q} ,  -{\bf Q}_\eta}), \\
\label{eq:D_q_infinite}
&{\bf D}_{\bf q} = 
 \sum_\eta {\bf D}_{{\bf Q}_\eta}  (\delta_{{\bf q} , {\bf Q}_\eta}-\delta_{{\bf q} , -{\bf Q}_\eta}).
\end{align} 
This corresponds to the model in the limit of infinite-range interactions in real space. 
The summations in Eqs.~\eqref{eq:J_q_infinite} and \eqref{eq:D_q_infinite} are taken for a particular set of the wave vectors ${\bf Q}_\eta$, which correspond to the nesting vectors of the Fermi surfaces when the model is constructed as an effective model for itinerant electron systems of the Kondo lattice type~\cite{Hayami2017,Hayami2018}. 
This infinite-range model was shown to stabilize VCs in two dimensions, which turn into SkXs in an applied magnetic field~\cite{Hayami2018}. 
It was also shown that the models with an additional infinite-range biquadratic interaction stabilize SkXs and HLs in two and three dimensions, respectively~\cite{Okumura2020,Shimizu2021a,Shimizu2021b}.

\subsection{Helical wave number and anisotropy}
\label{subsec:QJD}

In the present study, starting from the infinite-range model, we consider its extension by introducing exponential decay in the long-range interactions. 
Before going into the extension, we define the characteristic wave numbers ${\bf Q}_\eta$ and the anisotropy in the magnetic interaction in this section. 
With regard to ${\bf Q}_\eta$, for simplicity,
we take them being parallel to the principal axes of the hypercube and $|{\bf Q}_\eta|=Q$:
\begin{align}
{\bf Q}_\eta=
\begin{cases}
 Q\hat{\bf x}, & (d=1)\\
 Q\hat{\bf x}, Q\hat{\bf y}, & (d=2)\\
 Q\hat{\bf x}, Q\hat{\bf y}, Q\hat{\bf z}, & (d=3)
\end{cases},
\label{eq:Q_eta}
\end{align} 
where $\hat{\bf x}$, $\hat{\bf y}$, and $\hat{\bf z}$ are the unit vectors along the $x$, $y$, and $z$ axes, respectively.
Meanwhile, regarding the anisotropy, we introduce it in the symmetric part of the interaction $J_{\bf q}^{\alpha\alpha}$, following  Ref.~\cite{Hayami2018}. 
In the following, we describe the specific forms of the anisotropic interactions in each spatial dimension.

\subsubsection{One-dimensional case}\label{sec:inf_1D}
\begin{figure}[h]
  \centering
 \includegraphics[trim=0 370 0 0, clip,width= \columnwidth]{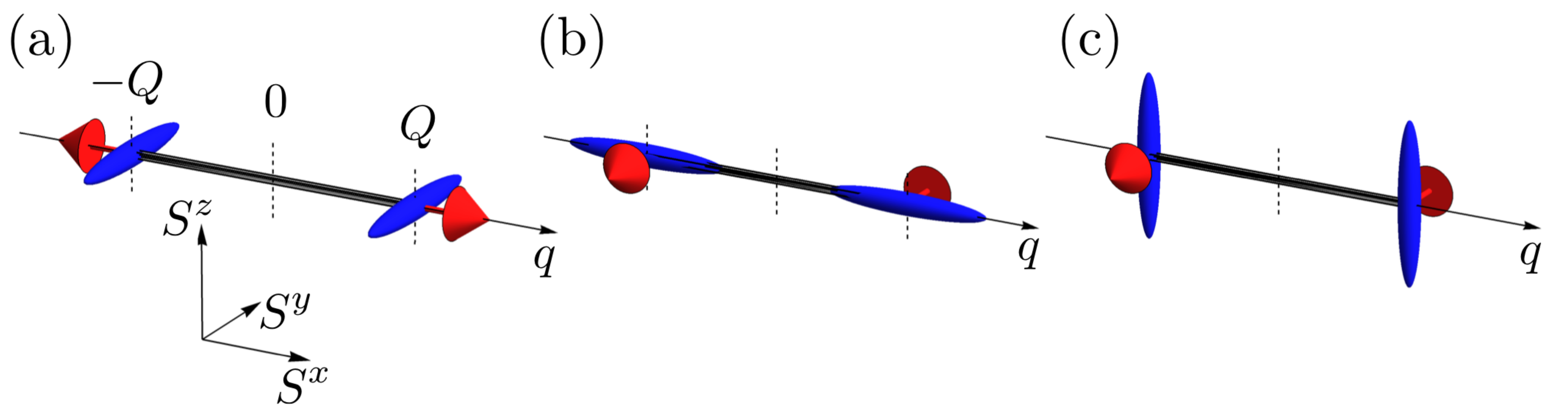}
  \caption{
  Pictorial representations of the coupling constants for the symmetric and antisymmetric interactions in the 1D models for 
  (a) proper-screw [Eqs.~\eqref{eq:Dq1d} and \eqref{eq:anisotropy}],
  (b) cycloid(I) [Eqs.~\eqref{eq:anisotropy2} and \eqref{eq:Dq1d2}], and
  (c) cycloid(II) [Eqs.~\eqref{eq:Dq1d2} and \eqref{eq:anisotropy3}] HS states.
  The blue ellipsoids represent $J^{\alpha\alpha}_{\pm Q}$: the lengths along the principal axes [100], [010], and [001] denote the amplitudes of $J^{xx}_{\pm Q}$, $J^{yy}_{\pm Q}$, and $J^{zz}_{\pm Q}$, respectively.
  The red arrows represent ${\bf D}_{\pm Q}$. 
  The axes for the spin space are shown in (a). 
 } \label{fig02}
\end{figure}
In the 1D case ($d=1$), we choose 
\begin{align}
&J^{\alpha \alpha}_q = J_{Q}^{\alpha\alpha} 
( \delta_{q , Q } +  \delta_{q , -Q }),
\label{eq:Jq1d}\\
&{\bf D}_q = {\bf D}_Q (\delta_{q, Q} - \delta_{q, -Q}),
\label{eq:Dq1d0}
\end{align}
with
\begin{align}
 Q  = \frac{2\pi}{\Lambda},
\end{align}
where $\Lambda$ gives the period of the HS states.
We consider three sets of the coupling constants with different spin anisotropy in $J_Q^{\alpha\alpha}$ and the direction of ${\bf D}_Q$ as described below. 
They are summarized in Table~\ref{tab1}, including the crystallographic point groups of the resultant models.

The first is the one which stabilizes a proper-screw HS state shown in Fig.~\ref{fig01}(a).  
In this case, to align the helical plane perpendicular to the propagating direction, we set ${\bf D}_{Q}$ as
\begin{align}
{\bf D}_Q = D \hat{\bf x}. 
\label{eq:Dq1d}
\end{align}
In addition, we introduce an anisotropy $\Delta$ in $J^{\alpha\alpha}_{Q}$ as
\begin{align}
J_{Q}^{xx} = J_{Q}^{zz} = J ( 1-\Delta),
\quad J_{Q}^{yy} =  J ( 1+2\Delta) .
\label{eq:anisotropy}
\end{align}
As we will discuss later, this anisotropy modulates
the spin helix from circular to elliptical and
makes the twist angle between neighboring spins nonuniform, which opens a gap in the magnetic excitation spectrum.
The pictorial representations of $J^{\alpha\alpha}_q$ and ${\bf D}_q$ are shown in Fig.~\ref{fig02}(a). 

The second one is for realizing a cycloidal HS state, whose spin structure is obtained by
$\pi/2$ spin rotation of the proper-screw one about the $z$ axis, as shown in Fig.~\ref{fig01}(b). 
To stabilize this, we rotate the spin axis in the coupling constants as
\begin{align}
&J_{Q}^{yy} = J_{Q}^{zz} = J ( 1-\Delta),
\quad J_{Q}^{xx} =  J ( 1+2\Delta), 
\label{eq:anisotropy2}\\
&{\bf D}_{Q} = D \hat{\bf y},
\label{eq:Dq1d2}
\end{align}
as shown in Fig.~\ref{fig02}(b). 
We call the spin state realized by this model the cycloid(I) HS state. 

The last one is for a different type of the cycloidal HS state, which we call cycloid(II), obtained by additional $\pi/2$ spin rotation about the $y$ axis. 
In this case, we set
\begin{align}
J_{Q}^{xx} = J_{Q}^{yy} = J ( 1-\Delta),
\quad J_{Q}^{zz} =  J ( 1+2\Delta), 
\label{eq:anisotropy3}
\end{align}
with the same ${\bf D}_{Q}$ as Eq.~\eqref{eq:Dq1d2}. 
This case is shown in Fig.~\ref{fig02}(c).

\subsubsection{Two-dimensional case}\label{sec:inf_2D}
\begin{figure}[h]
  \centering
 \includegraphics[trim=0 280 0 0, clip,width= \columnwidth]{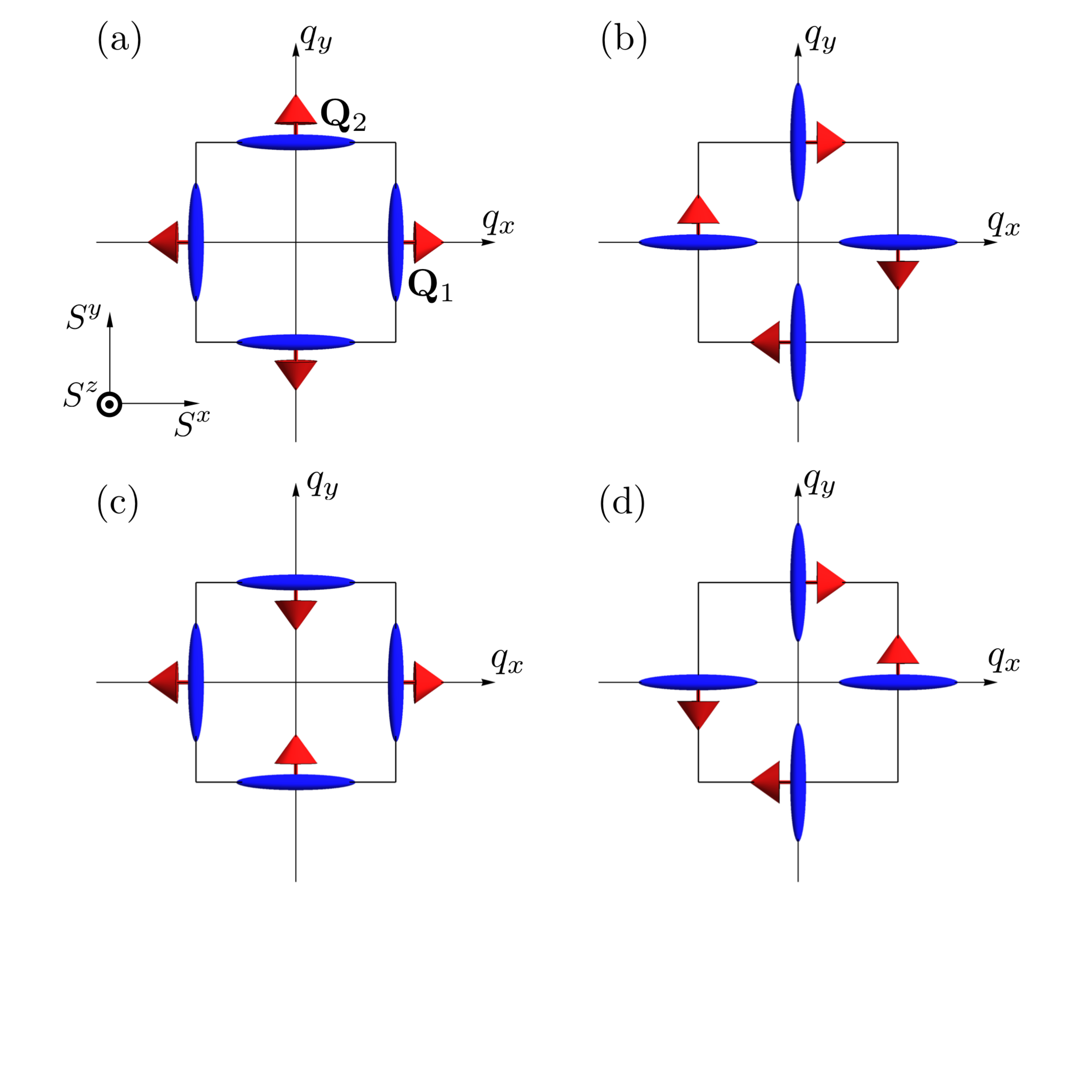}
  \caption{
   Similar pictorial representations to Fig.~\ref{fig02} for the 2D models realizing the VCs of
  (a) proper-screw(I) [Eqs.~\eqref{eq:DQeta2d} and \eqref{eq:JQeta2d1}], 
  (b) cycloid(I) [Eqs.~\eqref{eq:JQeta2d2} and \eqref{eq:DQeta2d2}], 
  (c) proper-screw(II) [Eqs.~\eqref{eq:JQeta2d3} and \eqref{eq:DQeta2d3}], and 
  (d) cycloid(II) [Eqs.~\eqref{eq:JQeta2d4} and \eqref{eq:DQeta2d4}] types.
  The notations are common to those in Fig.~\ref{fig02}. 
 } \label{fig03}
\end{figure}
In the 2D case ($d=2$), we choose 
\begin{align}
J^{\alpha\alpha}_{\bf q} =&  \sum_{\eta}  J^{\alpha\alpha}_{{\bf Q}_\eta}
(\delta_{{\bf q},{\bf Q}_\eta}+\delta_{{\bf q},-{\bf Q}_\eta}), \label{eq:Jq2d}\\
{\bf D}_{\bf q} =& \sum_{\eta}  {\bf D}_{{\bf Q}_\eta}
(\delta_{{\bf q}, {\bf Q}_\eta}-\delta_{{\bf q} ,-{\bf Q}_\eta}),\label{eq:Dq2d}
\end{align}
with
${\bf Q}_1=Q  \hat{\bf x}$ and ${\bf Q}_2=Q  \hat{\bf y}$ [see Eq.~\eqref{eq:Q_eta}].
We consider four sets of $J^{\alpha\alpha}_{{\bf Q}_\eta}$ and ${\bf D}_{{\bf Q}_\eta}$.
The first is the one which can stabilize a superposition of two proper-screw spirals. 
In this case, to align each helical plane perpendicular to the corresponding helical direction, we set ${\bf D}_{{\bf Q}_\eta}$ as
\begin{align}
{\bf D}_{{\bf Q}_\eta} =
\begin{cases}
 D \hat{\bf x},& (\eta=1)\\
 D \hat{\bf y},& (\eta=2)
\end{cases}.
\label{eq:DQeta2d}
\end{align}
For the symmetric part, we introduce the anisotropy compatible with $\mathsf{C}_4$ rotational or $\mathsf{S}_4$ rotoreflection symmetry about the $z$ axis, that is,
\begin{align}
&(J^{xx}_{{\bf Q}_\eta},J^{yy}_{{\bf Q}_\eta},J^{zz}_{{\bf Q}_\eta}) = \nonumber\\
&\begin{cases}
 [J ( 1-\Delta), J ( 1+2\Delta) , J ( 1-\Delta) ] , & (\eta=1)\\
 [J ( 1+2\Delta) ,J ( 1-\Delta), J ( 1-\Delta) ] , & (\eta=2)
\end{cases}.
\label{eq:JQeta2d1}
\end{align}
The pictorial representations of $J^{\alpha\alpha}_{{\bf q}}$ and ${\bf D}_{\bf q}$ are shown in Fig.~\ref{fig03}(a). 
We call the $2Q$ spin state stabilized in this setting the proper-screw(I) VC.

The second one is for realizing a superposition of two cycloidal spirals. 
Similar to the 1D case,
we apply $-\pi/2$ spin rotation to the first case about the $z$ axis and set
\begin{align}
&(J^{xx}_{{\bf Q}_\eta},J^{yy}_{{\bf Q}_\eta},J^{zz}_{{\bf Q}_\eta}) = \nonumber\\
&\begin{cases}
  [ J ( 1+2\Delta) ,J ( 1-\Delta), J ( 1-\Delta)  ] , & (\eta=1)\\
  [ J ( 1-\Delta), J ( 1+2\Delta) , J ( 1-\Delta) ], & (\eta=2)
\end{cases},
\label{eq:JQeta2d2}\\
&{\bf D}_{{\bf Q}_\eta} =
\begin{cases}
 - D \hat{\bf y} , & (\eta=1)\\
 \;\;\;  D \hat{\bf x} , & (\eta=2)
\end{cases}.
\label{eq:DQeta2d2}
\end{align}
See Fig.~\ref{fig03}(b). 
We call the 2$Q$ spin state stabilized in this setting the cycloid(I) VC. 

The third one is for realizing 
a superposition of two proper-screw spirals which are different from the first case.
This is obtained by additional $\pi$ spin rotation about the $[1\bar{1}0]$ axis,
and hence,
we set
\begin{align}
&(J^{xx}_{{\bf Q}_\eta},J^{yy}_{{\bf Q}_\eta},J^{zz}_{{\bf Q}_\eta}) = \nonumber\\
&\begin{cases}
 [J ( 1-\Delta), J ( 1+2\Delta) , J ( 1-\Delta) ] , & (\eta=1)\\
 [J ( 1+2\Delta) ,J ( 1-\Delta), J ( 1-\Delta) ] , & (\eta=2)
\end{cases},
\label{eq:JQeta2d3}\\
&{\bf D}_{{\bf Q}_\eta} =
\begin{cases}
 \;\;\;  D \hat{\bf x} , & (\eta=1)\\
  -D \hat{\bf y} , & (\eta=2)
\end{cases},
\label{eq:DQeta2d3}
\end{align}
as shown in Fig.~\ref{fig03}(c). 
We call the 2$Q$ spin state stabilized in this setting the proper-screw(II) VC. 

The last one is for realizing 
a different superposition of two cycloidal spirals from the second case.
This is obtained by additional $\pi/2$ spin rotation to the third case about the $z$ axis,
and hence,
we set
\begin{align}
&(J^{xx}_{{\bf Q}_\eta},J^{yy}_{{\bf Q}_\eta},J^{zz}_{{\bf Q}_\eta}) = \nonumber\\
&\begin{cases}
 [J ( 1+2\Delta) ,J ( 1-\Delta), J ( 1-\Delta) ] , & (\eta=1)\\
 [J ( 1-\Delta), J ( 1+2\Delta) , J ( 1-\Delta) ] , & (\eta=2)
\end{cases},
\label{eq:JQeta2d4}\\
&{\bf D}_{{\bf Q}_\eta} =
\begin{cases}
   D \hat{\bf y} , & (\eta=1)\\
   D \hat{\bf x} , & (\eta=2)
\end{cases},
\label{eq:DQeta2d4}
\end{align}
as shown in Fig.~\ref{fig03}(d). 
We call the 2$Q$ spin state stabilized in this setting the cycloid(II) VC. 

The four sets of the coupling constants are summarized in Table~\ref{tab1}, including the crystallographic point groups of the resultant models.
See also Fig.~\ref{fig15} for the spin configurations of each $2Q$ spin state. 
We note that the proper-screw(I) VC is categorized into the so-called Bloch-type VCs, while the cycloid(I) VC is the so-called N\'eel-type. 
In general, 
the Bloch- and N\'eel-type multiple-$Q$ spin states are realized 
under the Rashba- and Dresselhaus-type spin-orbit couplings, respectively~\cite{Rowland2016,Hayami2018}.
Note that multiple-$Q$ states in the presence of both types of the spin-orbit coupling were studied for a model which explicitly includes itinerant electrons~\cite{Okada2018}.

\subsubsection{Three-dimensional case}\label{sec:inf_3D}
\begin{figure}[h]
  \centering
 \includegraphics[trim=0 120 0 0, clip,width= \columnwidth]{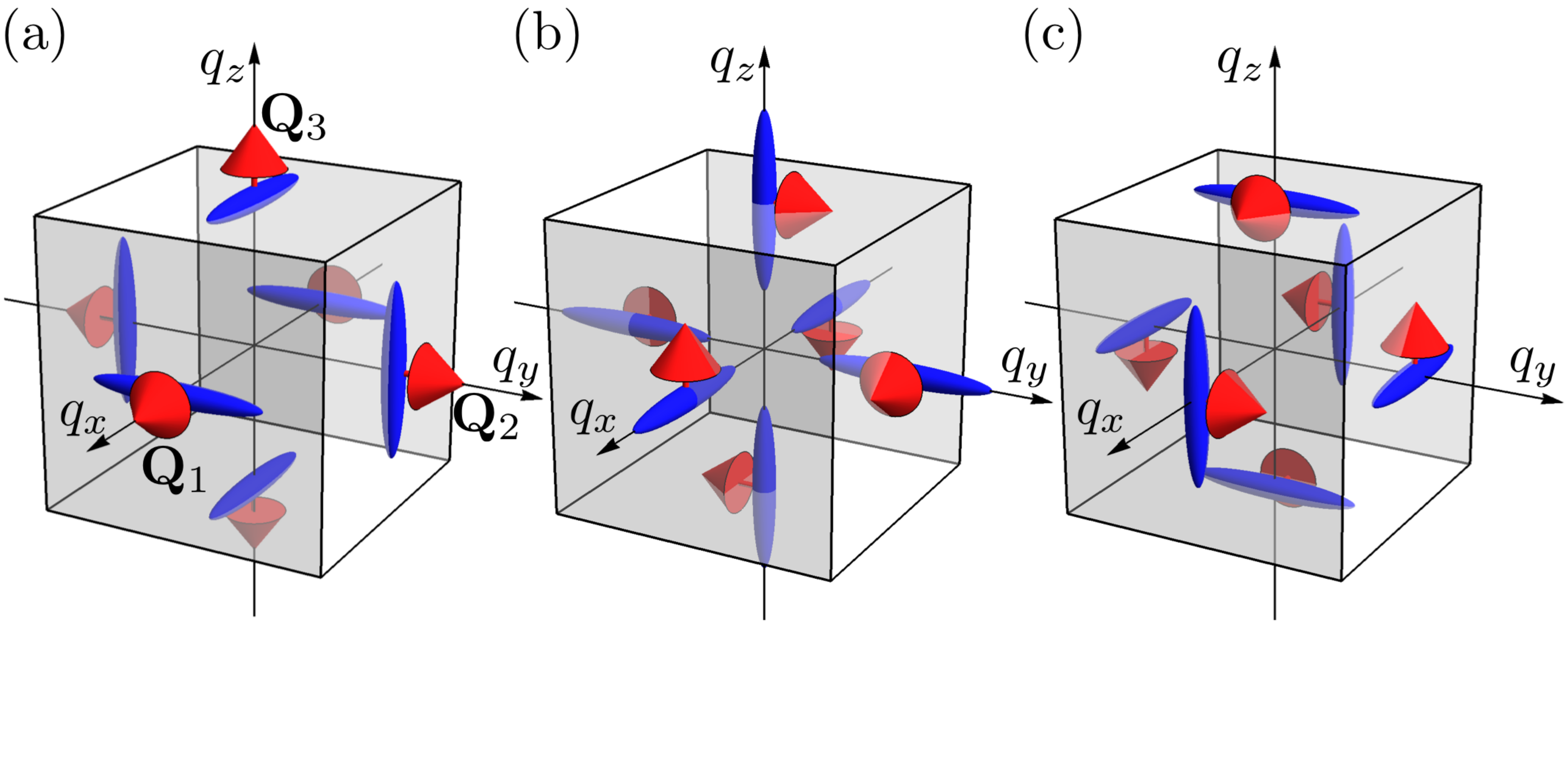}
  \caption{
   Similar pictorial representations to Fig.~\ref{fig02} for the 3D models realizing the HLs of
   (a) proper-screw [Eqs.~\eqref{eq:DQeta3d} and \eqref{eq:JQeta3d1}],
   (b) cycloid(I) [Eqs.~\eqref{eq:JQeta3d2} and \eqref{eq:DQeta3d2}], and
   (c) cycloid(II) [Eqs.~\eqref{eq:JQeta3d3} and \eqref{eq:DQeta3d3}] types.
   The notations are common to those in Fig.~\ref{fig02}.   
 } \label{fig04}
\end{figure}
In the 3D case ($d=3$), we choose the same forms of $J^{\alpha\alpha}_{\bf q}$ and ${\bf D}_{\bf q}$
as Eqs.~\eqref{eq:Jq2d} and \eqref{eq:Dq2d},
but with ${\bf Q}_1=Q  \hat{\bf x}$, ${\bf Q}_2=Q  \hat{\bf y}$, and ${\bf Q}_3=Q  \hat{\bf z}$ [see Eq.~\eqref{eq:Q_eta}]. 
We consider three sets of $J^{\alpha\alpha}_{{\bf Q}_\eta}$ and ${\bf D}_{{\bf Q}_\eta}$ as described below; see Table~\ref{tab1}.
See also Fig.~\ref{fig19} for the spin configurations stabilized in each model.

The first is the one which can stabilize a superposition of three proper-screw spirals.
In this case, to align the helical plane perpendicular to the corresponding helical direction, we set ${\bf D}_{\bf Q}$ as
\begin{align}
{\bf D}_{{\bf Q}_\eta} =
\begin{cases}
 D \hat{\bf x} , & (\eta=1)\\
 D \hat{\bf y} , & (\eta=2)\\
 D \hat{\bf z} , & (\eta=3)
\end{cases}.
\label{eq:DQeta3d}
\end{align}
For the symmetric part, we introduce the anisotropy compatible with
$\mathsf{C}_3$ rotational symmetry about the $[111]$ axis, that is,
\begin{align}
&(J^{xx}_{{\bf Q}_\eta},J^{yy}_{{\bf Q}_\eta},J^{zz}_{{\bf Q}_\eta}) = \nonumber\\
&\begin{cases}
 [J ( 1-\Delta), J ( 1+2\Delta) , J ( 1-\Delta) ] , & (\eta=1)\\
 [J ( 1-\Delta), J ( 1-\Delta) , J ( 1+2\Delta) ] , & (\eta=2)\\
 [J ( 1+2\Delta) ,J ( 1-\Delta), J ( 1-\Delta) ] , & (\eta=3)
\end{cases}.
\label{eq:JQeta3d1}
\end{align}
The pictorial representations of $J^{\alpha\alpha}_{{\bf q}}$ and ${\bf D}_{\bf q}$ are shown in Fig.~\ref{fig04}(a). 
We call the 3$Q$ spin state realized in this setting the proper-screw HL.

The second one is for realizing a superposition of three cycloidal spirals.
This is obtained by $-2\pi/3$ spin rotation of the proper-screw HL about the $[111]$ axis,
and hence,
we set
\begin{align}
&(J^{xx}_{{\bf Q}_\eta},J^{yy}_{{\bf Q}_\eta},J^{zz}_{{\bf Q}_\eta}) = \nonumber\\
&\begin{cases}
 [J ( 1+2\Delta) ,J ( 1-\Delta), J ( 1-\Delta) ] , & (\eta=1)\\
 [J ( 1-\Delta), J ( 1+2\Delta) , J ( 1-\Delta) ] , & (\eta=2)\\
 [J ( 1-\Delta), J ( 1-\Delta) , J ( 1+2\Delta) ] , & (\eta=3)
\end{cases},
\label{eq:JQeta3d2}\\
&{\bf D}_{{\bf Q}_\eta} =
\begin{cases}
D \hat{\bf z} , & (\eta=1)\\
 D \hat{\bf x} , & (\eta=2)\\
 D \hat{\bf y} , & (\eta=3)
\end{cases}.
\label{eq:DQeta3d2}
\end{align}
See Fig.~\ref{fig04}(b). 
We call the $3Q$ spin state stabilized in this setting the cycloid(I) HL. 

The last one is for realizing a different superposition of three cycloidal spirals.
This is obtained by additional $-2\pi/3$ spin rotation about the $[111]$ axis,
and hence,
we set
\begin{align}
&(J^{xx}_{{\bf Q}_\eta},J^{yy}_{{\bf Q}_\eta},J^{zz}_{{\bf Q}_\eta}) = \nonumber\\
&\begin{cases}
 [J ( 1-\Delta), J ( 1-\Delta) , J ( 1+2\Delta) ] , & (\eta=1)\\
 [J ( 1+2\Delta) ,J ( 1-\Delta), J ( 1-\Delta) ] , & (\eta=2)\\
 [J ( 1-\Delta), J ( 1+2\Delta) , J ( 1-\Delta) ] , & (\eta=3)
\end{cases},
\label{eq:JQeta3d3}\\
&{\bf D}_{{\bf Q}_\eta} =
\begin{cases}
 D \hat{\bf y} , & (\eta=1)\\
D \hat{\bf z} , & (\eta=2)\\
 D \hat{\bf x} , & (\eta=3)\\
\end{cases},
\label{eq:DQeta3d3}
\end{align}
as shown in Fig.~\ref{fig04}(c). 
We call the 3$Q$ spin state stabilized in this setting the cycloid(II) HL.

\subsection{Finite-range model}
\label{subsec:Finite-range model}
As introduced in Sec.~\ref{subsec:Infinite-range limit}, the model in Eq.~\eqref{eq:H} has been studied in the limit of the infinite-range interactions in Eqs.~\eqref{eq:J_q_infinite} and \eqref{eq:D_q_infinite}. 
In the following, we extend the model by introducing spatial decay in the interactions.
After explaining the extension in detail for the 1D case in Sec.~\ref{subsubsec:finite-range 1D}, we describe the 2D and 3D cases in Secs.~\ref{subsubsec:finite-range 2D} and \ref{subsubsec:finite-range 3D}, respectively. 
Throughout this section, we assume the set of the coupling constants for the proper-screw states firstly introduced for each dimensional case in Sec.~\ref{subsec:QJD}; 
the extensions to the other sets are straightforward by using the spin rotations introduced above.

\subsubsection{One-dimensional case}
\label{subsubsec:finite-range 1D}
\begin{figure}[H]
  \centering
  \includegraphics[trim=0 0 0 0, clip,width= \columnwidth]{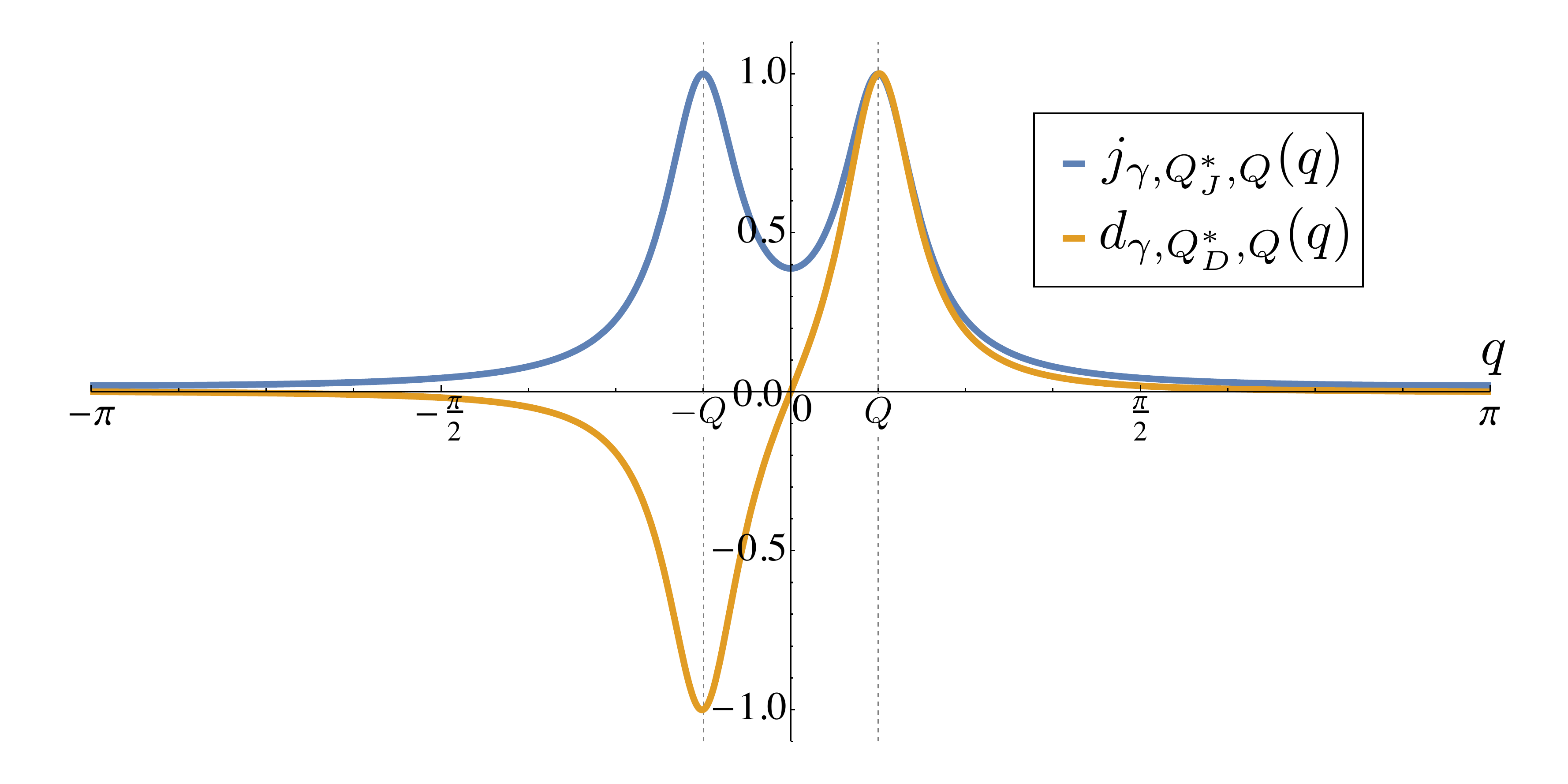}
  \caption{
  $q$ dependences of $j_{\gamma,Q^*_J,Q}(q)$ [Eq.~\eqref{eq:jQq}] and $d_{\gamma,Q^*_D,Q}(q)$ [Eq.~\eqref{eq:dQq}]
  in the 1D finite-range model. 
  We take $\Lambda = 16$ and $\gamma=0.2$.
  $Q_J^*$ and $Q_D^*$ are set to $Q_J^* \simeq 0.396$ and $Q_D^* \simeq 0.390$ so that $J^{\alpha\alpha}_q$ and $|D_q|$ take their maxima at $q =  \pm Q  =  \pm \pi/8$.
 } \label{fig05}
\end{figure}
Let us begin with the real-space representation of the 1D infinite-range model with Eqs.~\eqref{eq:Jq1d} and \eqref{eq:Dq1d0}.
By the Fourier transformation, the Hamiltonian reads
\begin{align}
\mathcal{H} 
= -\frac{2}{L} \sum_{\ell , \ell_1}
\sum_{\alpha,\beta}
\mathcal{J}_{ Q }^{\alpha\beta}
S^{\alpha}_\ell 
S^{\beta}_{\ell-\ell_1}
e^{- i Q \ell_1},
\end{align}
where the sum of $\ell_1$ runs over all the integers in the range $[-L/2, L/2)$. 
Here and hereafter, we assume that the helix has a commensurate period to the lattice for simplicity; namely, $L/\Lambda$ is an integer.
To introduce spatial decay in the infinite-range interactions, we multiply an exponential dumping factor as
\begin{align}
\mathcal{H} \to
\tilde{\mathcal{H}}
= -\frac{2}{L} \sum_{\ell , \ell_1}
\sum_{\alpha,\beta}
\mathcal{J}_{ Q }^{\alpha\beta}
S^{\alpha}_\ell 
S^{\beta}_{\ell-\ell_1}
e^{- i Q \ell_1}
e^{- \gamma | \ell_1 |} .
\end{align}
For sufficiently large $L$, the modified Hamiltonian $\tilde{\mathcal{H}}$ can be expressed as
\begin{align}
\tilde{\mathcal{H}} =&
-\frac{2}{L} \sum_{q \in {\rm 1BZ}}
\sum_{\alpha,\beta}
\mathcal{J}_{ Q }^{\alpha\beta}
S^{\alpha}_q 
S^{\beta}_{-q}
f_{\gamma , Q }(q) ,
\end{align}
where
\begin{align}
f_{\gamma,Q} (q)
=
\frac{\sinh \gamma}{\cosh \gamma - \cos(Q -q)}.
\end{align}
This function $f_{\gamma,Q} (q)$ for $\gamma \ll 1$ is well approximated near $q = Q$ by the Lorentzian function as
\begin{align}
f_{\gamma,Q} (q) \approx \frac{ 2 \gamma}{ \gamma^2 + (Q-q)^2 }.
\end{align}
By symmetrizing the terms of $\pm q$, we end up with the Hamiltonian in the form
\begin{align}
\tilde{\mathcal{H}} =&
- \sum_{q \in {\rm 1BZ}}
\sum_{\alpha,\beta}
\tilde{\mathcal{J}}_{q}^{\alpha\beta}
S^{\alpha}_q 
S^{\beta}_{-q},
\label{eq:H_tilde_1D}
\end{align}
where 
\begin{align}
\tilde{\mathcal{J}}_q^{\alpha\beta}=
\begin{cases}
\frac{\mathcal{J}_{Q}^{\alpha\alpha}}L
\left[ f_{\gamma , Q }(q) + f_{\gamma , Q }(-q) \right]
 ,&(\alpha = \beta)\\
\frac{\mathcal{J}_{Q}^{\alpha\beta}}L
\left[ f_{\gamma , Q }(q) - f_{\gamma , Q }(-q) \right]
,&(\alpha \neq \beta)\\
\end{cases}.
\end{align}

The model in Eq.~\eqref{eq:H_tilde_1D} stabilizes a spin helix whose period deviates from $\Lambda = 2\pi/Q$ because the peaks of $|\tilde{\mathcal{J}_q^{\alpha\beta}}|$ are shifted due to the factors of $f_{\gamma , Q }(\pm q)$.
To facilitate the following analyses, we adjust the form of the interactions so that $|\tilde{\mathcal{J}}_q^{\alpha\beta}|$ have peaks exactly at $q=\pm  Q $ and the period of the stable spin helix becomes $\Lambda = 2\pi/Q$. 
This is achieved by replacing $f_{\gamma , Q }(\pm q)$ by $f_{\gamma , Q^*}(\pm q)$, where $Q^*$ is determined so that the derivative of the corresponding coupling constant with respect to $q$ becomes zero at $q=\pm Q$. 
In addition, we rescale all the elements of $\tilde{\mathcal{J}}_{q}^{\alpha\beta}$ individually so that they take the same values with the infinite-range model at $q=\pm  Q $, namely, $\tilde{\mathcal{J}}_Q^{\alpha\beta} = \mathcal{J}_Q^{\alpha\beta}$.
Then, finally we obtain the Hamiltonian with the finite-range interactions in the same form of Eqs.~\eqref{eq:H} and \eqref{eq:Hq} with
\begin{align}
& J^{\alpha\alpha}_q = J_{Q}^{\alpha\alpha} j_{\gamma, Q_J^*, Q } (q), \label{eq:frJq1d} \\
& {\bf D}_q = {\bf D}_Q d_{\gamma, Q_D^*, Q } (q) 
\label{eq:frDq1d},
\end{align}
where $J^{\alpha\alpha}_Q$ and ${\bf D}_Q$ are given in Eqs.~\eqref{eq:anisotropy} and \eqref{eq:Dq1d}, respectively, for the proper-screw HS case;
$j_{\gamma, Q_J^*, Q }$ and $d_{\gamma, Q_J^*, Q }$ are defined as
\begin{align}
&j_{\gamma, Q_J^*, Q } (q) =
\frac{
f_{\gamma,Q^*_J}(q)
+
f_{\gamma,Q^*_J}(-q)
}{
f_{\gamma,Q^*_J}( Q )
+
f_{\gamma,Q^*_J}(- Q )
}, \label{eq:jQq} \\
&d_{\gamma, Q_D^*, Q } (q) =
\frac{
f_{\gamma,Q^*_D}(q)
-
f_{\gamma,Q^*_D}(-q)
}{
f_{\gamma,Q^*_D}( Q )
-
f_{\gamma,Q^*_D}(- Q )
}.  \label{eq:dQq}
\end{align}
Here, $Q^*_J$ and $Q^*_D$ are determined by
solving
\begin{align}
&\left.\frac{\partial j_{\gamma, Q_J^*, Q } (q)}{\partial q} \right|_{q= Q } 
=0,\\
&\left.\frac{\partial  d_{\gamma, Q_D^*, Q } (q)  }{\partial q} \right|_{q= Q } 
 =0,
\end{align}
respectively.
Figure~\ref{fig05} exemplifies $j_{\gamma, Q_J^*, Q }(q)$ and $d_{\gamma, Q_D^*, Q }(q)$ for $\Lambda = 16$ and $\gamma=0.2$.
The other cases for the cycloidal HS states are obtained by the spin rotations in Sec.~\ref{sec:inf_1D}.

\subsubsection{Two-dimensional case}
\label{subsubsec:finite-range 2D}
\begin{figure}[H]
  \centering
   \includegraphics[trim=0 0 0 0, clip,width= \columnwidth]{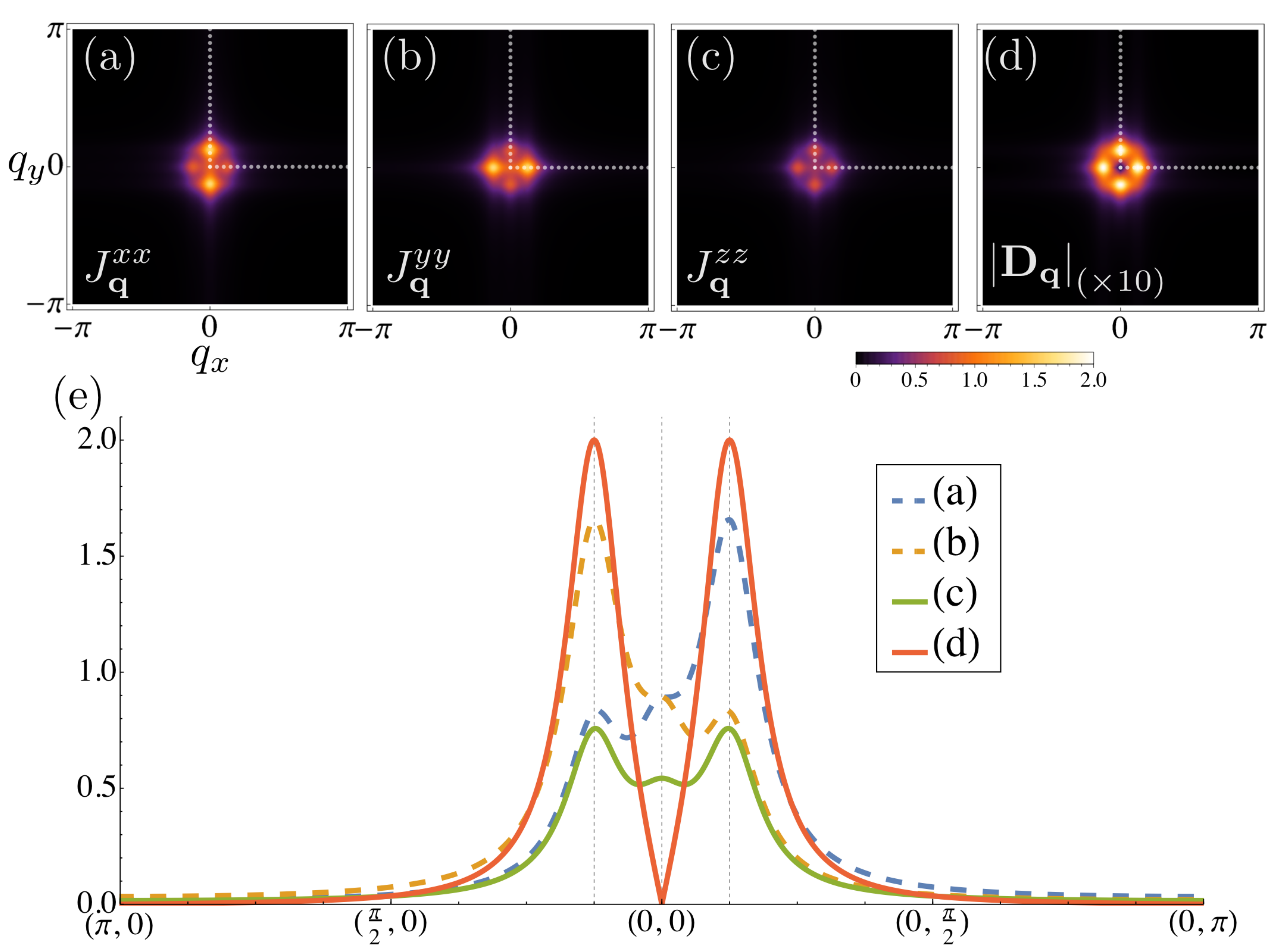}
  \caption{${\bf q}=(q_x, q_y)$ dependences of
  (a) $J^{xx}_{\bf q}$, 
  (b) $J^{yy}_{\bf q}$, 
  (c) $J^{zz}_{\bf q}$, 
  and (d) $|{\bf D}_{\bf q}|$ in the 2D finite-range model. 
  The color bar is common to 
  (a--d), while $|{\bf D}_{\bf q}|$ in (d) is plotted by multiplying a factor of 10 for better visibility.
  (e) Profile of (a--d) along the path $(\pi,0)$--$(0,0)$--$(0,\pi)$ [white dotted lines in (a--d)]. 
  We take $\Lambda = 16$, $\gamma=0.2$, $J=1$, $D=0.2$, and $\Delta = 0.3$.
  The values for $Q_J^*$ and $Q_D^*$ are the same as those in Fig.~\ref{fig05}.
 } \label{fig06}
\end{figure}
Following the 1D case, we can construct the finite-range model in two dimensions.
The Hamiltonian also has the same form of Eqs.~\eqref{eq:H} and \eqref{eq:Hq}.
Using the functions $j_{\gamma, Q_J^*, Q } (q)$ and $d_{\gamma, Q_D^*, Q } (q) $
in Eqs.~\eqref{eq:jQq} and \eqref{eq:dQq}, respectively, the coupling constants for the symmetric interactions are given as
\begin{align}
{J}^{\alpha\alpha}_{\bf q} &= 
J^{\alpha\alpha}_{{\bf Q}_1} 
j_{\gamma, Q ,Q_J^*}(q_x) ~ j_{\gamma,0,0}(q_y)\nonumber\\
&+
J^{\alpha\alpha}_{{\bf Q}_2} 
j_{\gamma,0,0}(q_x) ~ j_{\gamma, Q ,Q_J^*}(q_y),
%
\end{align}
with $J^{\alpha\alpha}_{{\bf Q}_\eta}$ in Eq.~\eqref{eq:JQeta2d1}, 
and those for the antisymmetric interactions are given as
\begin{align}
{\bf D}_{\bf q} =
D_{\bf q} \frac{{\bf q}}{|{\bf q}|}, 
\label{eq:D_q_1D_finite-range}
\end{align}
where
\begin{align}
D_{\bf q}&=D\left[
\left|
d_{\gamma,  Q ,Q_D^*}(q_x)
j_{\gamma, 0,0}(q_y)
\right| \right. \nonumber\\
&~\;\;\;\;+\left. 
\left|
j_{\gamma, 0,0}(q_x)
d_{\gamma,  Q ,Q_D^*}(q_y)
\right|
\right],
\end{align}
for the case of the proper-screw(I) VC. 
Figure~\ref{fig06} exemplifies $J_{\bf q}^{\alpha\alpha}$ and $D_{\bf q}$ for $\Lambda = 16$, $\gamma=0.2$, $J=1$, $D=0.2$, and $\Delta = 0.3$.
The other cases are obtained by the proper spin rotations in Sec.~\ref{sec:inf_2D}.

\subsubsection{Three-dimensional case}
\label{subsubsec:finite-range 3D}
\begin{figure}[h]
  \centering
   \includegraphics[trim=0 270 0 0, clip,width= \columnwidth]{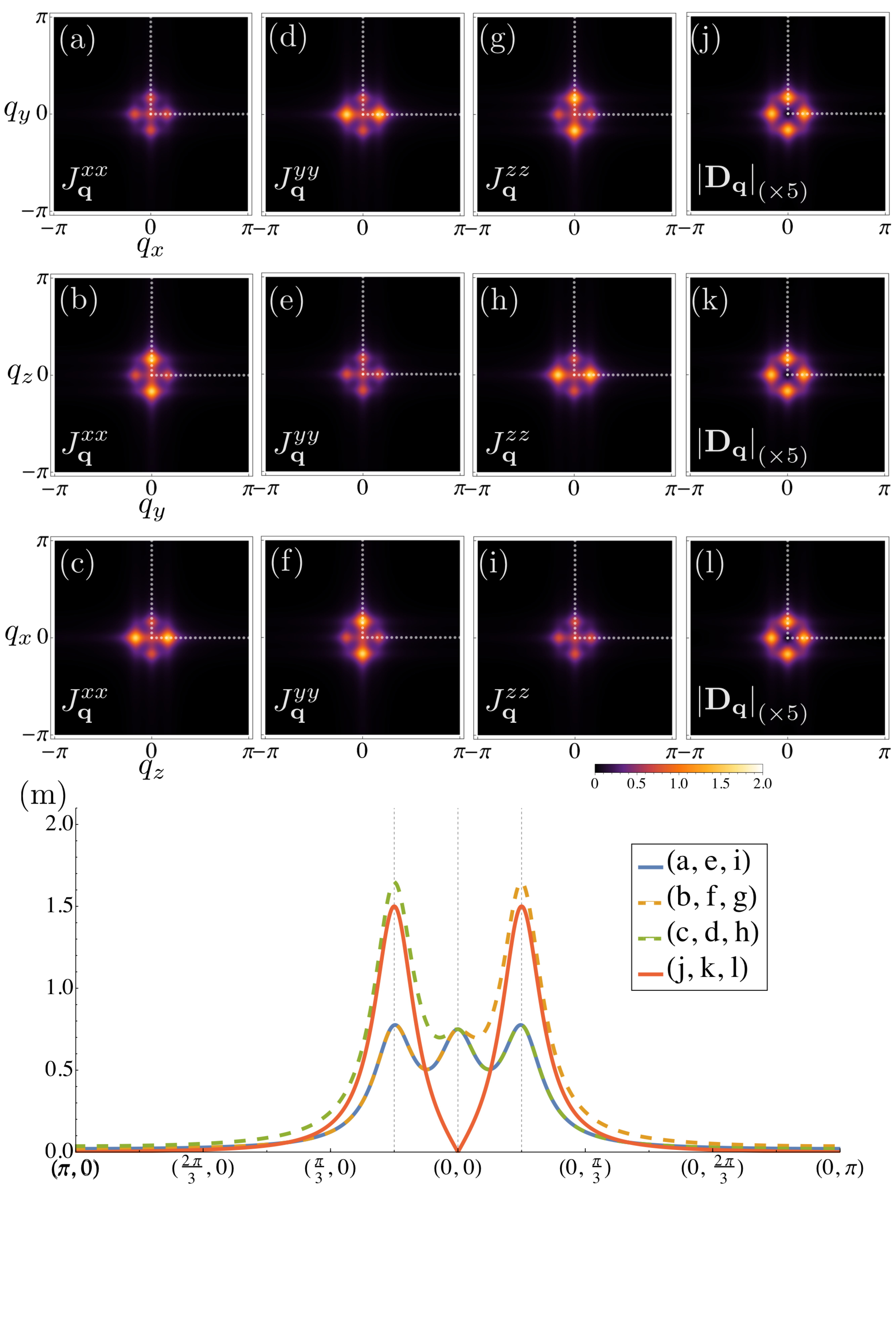}
  \caption{${\bf q}=(q_x,q_y,q_z)$ dependences of
  [(a)--(c)] $J^{xx}_{\bf q}$, 
  [(d)--(f)] $J^{yy}_{\bf q}$, 
  [(g)--(i)] $J^{zz}_{\bf q}$, 
   and [(j)--(l)] $|{\bf D}_{\bf q}|$ in the 3D finite-range model. 
   (a,d,g,j), (b,e,h,k), and (c,f,i,l) show the $q_z=0$, $q_x=0$, and $q_y=0$ planes, respectively.
   The color bar is common to (a--l), while $|{\bf D}_{\bf q}|$ in (j,k,l) are plotted by multiplying a factor of 5 for better visibility.
  (m) Profile along the white dotted lines in (a--l). 
  We take $\Lambda = 12$, $\gamma=0.2$, $J=1$, $D=0.3$, and $\Delta = 0.3$. 
   $Q_J^*$ and $Q_D^*$ are set to $Q_J^* = 0.525$ and $Q_D^* = 0.522$ 
   so that $J^{\alpha\alpha}_q$ and $|\bf{D}_{\bf q}|$ take their maxima at $q = \pm Q= \pm \pi/6$.
 } \label{fig07}
\end{figure}
In a similar manner, we can obtain the forms of the finite-range interactions for the 3D case as
\begin{align}
J^{\alpha\alpha}_{\bf q} 
&=
J^{\alpha\alpha}_{{\bf Q}_1} 
j_{\gamma, Q ,Q_J^*}(q_x) ~ j_{\gamma,0,0}(q_y) ~ j_{\gamma,0,0}(q_z) \nonumber\\
&+
J^{\alpha\alpha}_{{\bf Q}_2} 
j_{\gamma,0,0}(q_x) ~ j_{\gamma, Q ,Q_J^*}(q_y) ~ j_{\gamma,0,0}(q_z) \nonumber\\
&+
J^{\alpha\alpha}_{{\bf Q}_3} 
j_{\gamma,0,0}(q_x)  ~ j_{\gamma,0,0}(q_y) ~ j_{\gamma, Q ,Q_J^*}(q_z),\\
%
D_{\bf q} &= D \left[
\left|
d_{\gamma,  Q ,Q_D^*}(q_x)
j_{\gamma, 0,0}(q_y)
j_{\gamma, 0,0}(q_z)
\right| \right. \nonumber\\
&\,\;\;\;\;\;\left.+
\left|
j_{\gamma, 0,0}(q_x)
d_{\gamma,  Q ,Q_D^*}(q_y)
j_{\gamma, 0,0}(q_z)
\right| \right.\nonumber\\
&\,\;\;\;\;\;\left.+
\left|
j_{\gamma, 0,0}(q_x)
j_{\gamma, 0,0}(q_y)
d_{\gamma,  Q ,Q_D^*}(q_z)
\right|
\right],
\end{align}
where we use Eqs.~\eqref{eq:JQeta3d1} and \eqref{eq:D_q_1D_finite-range} for $J^{\alpha\alpha}_{{\bf Q}_\eta}$ and $D_{\bf q}$, respectively, in the case of the proper-screw HL.
Figure~\ref{fig07} exemplifies $J_{\bf q}^{\alpha\alpha}$ and $D_{\bf q}$ for $\Lambda = 12$, $\gamma=0.2$, $J=1$,
$D=0.3$, and $\Delta = 0.3$. 
The other cases are obtained by the proper spin rotations in Sec.~\ref{sec:inf_3D}.

\section{Methods}\label{sec:method}
\subsection{Variational method}\label{sec:variational}
In this study, we investigate the spin excitation spectrum of the stable ground state for each model introduced in the previous section. 
For this purpose, we first determine the ground state by using variational calculations
in the classical limit where ${\bf S}_{\bf r}$ is regarded as a 3D vector with fixed length of $|{\bf S}_{\bf r}| = 1$. 
In the case of the isotropic symmetric interactions ($\Delta=0$), 
we perform the variational calculation analytically by assuming a $1Q$ HS state with a uniform twist in  
all dimensions, as we do not find any other lower-energy state in the numerical variational calculation described below. 
Meanwhile, in the presence of the spin anisotropy with $\Delta \neq 0$, 
we employ the numerical variational calculation, as the $1Q$ HS state is modulated and other multiple-$Q$ spin states may have lower energy. 
In the numerical calculation, 
starting from several different initial spin configurations (see below), we determine the lowest-energy state by optimization of the individual spin orientation taking into account 
the internal magnetic field from the other spins and the single-ion anisotropy $(S^\alpha_{\bf r})^2$ appearing in the real-space form of the Hamiltonian.
As the initial spin configurations, we take into account a $1Q$ HS state with a uniform twist for the 1D case, the $1Q$ state and a 2$Q$ VC~\cite{Hayami2018} for the 2D case, 
and the $1Q$ and $2Q$ states and a 3$Q$ HL~\cite{Okumura2020} for the 3D case; in each state, we set an
appropriate helical plane depending on the type of ${\bf D}_{\bf q}$, namely, the proper-screw type (${\bf D}_{\bf q} \parallel {\bf q}$) or cycloid type (${\bf D}_{\bf q} \perp {\bf q}$).
We first perform the numerical calculations for
the infinite-range model with
$\gamma=0$, and then, study the finite-range model with $\gamma>0$
starting from the solution for the infinite-range model as the initial state.

\subsection{Linear spin-wave theory}\label{sec:LSWtheory}
For the stable spin configuration obtained by the variational method, we study the spin excitation by using the linear spin-wave theory. 
For the 1D $1Q$ HS states with uniform twists, we obtain the analytic form of the excitation spectra regardless of the range of interactions (Sec.~\ref{sec:analytic1d}).
Meanwhile, for the anisotropic cases ($\Delta > 0$) as well as the 2D and 3D cases, we perform the spin-wave calculations numerically as follows. 
For each stable spin configuration, we introduce new local spin axes at each site so that all the spins point to the $z$ direction. 
We denote the spins in the new spin frame as 
$\tilde{\bf S}_{\bf r}$. 
Then, the stable spin configuration is regarded as a ferromagnetic state, namely, $\tilde{\bf S}_{\bf r} = \hat{\bf z}$ for all ${\bf r}$. 
We apply the Holstein-Primakoff transformation 
to the Hamiltonian in the new spin frame, 
leaving the lowest order of bosonic operators:
\begin{align}
\begin{bmatrix}
\tilde{S}_{\bf r}^x\\
\tilde{S}_{\bf r}^y\\
\tilde{S}_{\bf r}^z\\
\end{bmatrix}
\to
\begin{bmatrix}
\sqrt{\frac{S}{2}}(a^{\;}_{\bf r} + a^\dag_{\bf r} ) \\
\sqrt{\frac{S}{2}}\frac{1}{{i}} (a^{\;}_{\bf r} - a^\dag_{\bf r})\\
S -  a^\dag_{\bf r} a^{\;}_{\bf r} \\
\end{bmatrix},
\label{eq:HPtrans}
\end{align}
where $a_{\bf r}$ and $a^\dag_{\bf r}$ represent the annihilation and creation operators of magnon at site ${\bf r}$, respectively; 
$S$ is the spin quantum number of ${\bf S}_{\bf r}$. 

We denote the spatial coordinate ${\bf r}$ as ${\bf r}={\bf R} + {\bf r}_0$, 
where ${\bf R}$ and ${\bf r}_0$ are the position vectors of each magnetic unit cell and the sublattice site 
within the unit cell, respectively:
$\{ {\bf R}  | R^\mu = \Lambda N^\mu , N^\mu \in [0, L/\Lambda) \}$ 
and $\{ {\bf r}_0  | r^\mu_0 \in [0, \Lambda) \}$, where $N^\mu$ and $r^\mu_0$ are integers. 
The Brillouin zone is folded from $\{ {\bf q} | -\pi \leq q^\mu < \pi \}$ to $\{ {\bf K} | -\pi/\Lambda \leq K^\mu < \pi/\Lambda \} $ under the magnetic order with period of $\Lambda$.
Using the Fourier transformation
\begin{align}
a_{{\bf K},{\bf r}_{0}}=\left(\frac{\Lambda}{L}\right)^d \sum_{\bf R} a_{{\bf R}+{\bf r}_{0}}
e^{+i {\bf K}\cdot{\bf R}} ,
\end{align}
we obtain the linear spin-wave Hamiltonian expressed as
\begin{align}
\mathcal{H}_{\rm SW}=
\frac{S}{2 \Lambda^d}
 { \sum_{\bf K} }'
{\bm \alpha}^\dag_{\bf K}
\mathsf{A}^{\;}_{{\bf K}}
{\bm \alpha}^{\;}_{\bf K},
\label{eq:H_SW}
\end{align}
where
\begin{align}
{\bm \alpha}^\dag_{{\bf K}} =&
\left[
{\bf a}_{{\bf K}+} ,
~
{\bf a}_{-{\bf K}+} ,
~
{\bf a}_{{\bf K}-} ,
~
{\bf a}_{-{\bf K}-}
\right],
\end{align}
with
\begin{align}
{\bf a}_{{\bf K}+}=&
\left[
a^\dag_{{\bf K},(0,\cdots)} ,
\cdots,
a^\dag_{{\bf K},{\bf r}},
\cdots ,
a^\dag_{{\bf K},(\Lambda-1,\cdots)}
\right],\\
{\bf a}_{{\bf K}-}=&
\left[
a^{\;}_{{\bf K},(0,\cdots)}
 ,
 \cdots
  ,
a^{\;}_{{\bf K},{\bf r}}
 ,
 \cdots
 ,
a^{\;}_{{\bf K},(\Lambda-1,\cdots)}
\right]. 
\end{align}
In Eq.~\eqref{eq:H_SW}, $\mathsf{A}_{\bf K}$ is a $4\Lambda^d \times 4\Lambda^d$ matrix for generic ${\bf K}$, while it becomes
a $2\Lambda^d \times 2\Lambda^d$ matrix 
for ${\bf K}=0$ or on the zone boundary;
each term in the sum of ${\bf K}$ includes all the contributions from $a^{\;}_{\pm {\bf K},{\bf r}_{0}}$ and $a^\dag_{\pm {\bf K},{\bf r}_{0}}$, 
and the sum $\sum'_{\bf K}$ runs over a half of the folded Brillouin zone (e.g., $K^x \geq 0$). 
By the Bogoliubov transformation~\cite{Colpa1978}, the Hamiltonian is diagonalized as 
\begin{align}
\mathcal{H}_{\rm SW}=
{ \sum_{\bf K} }'
\sum_{p}
\varepsilon_{{\bf K}p}
b^\dag_{{\bf K}p}
b^{\;}_{{\bf K}p}
+{\rm const}.,
\label{eq:HSWdiag}
\end{align}
where 
$\varepsilon_{{\bf K}p} > 0$ represents the $p$th spin-wave dispersion ($p=1,\cdots,  \dim [\mathsf{A}_{\bf K}]/2$), 
and $b^{\;}_{{\bf K}p}$ and $b^\dag_{{\bf K}p}$ represent the annihilation and creation operators of a bosonic quasiparticle, respectively, 
which are given by linear combinations of $a^{\;}_{\pm{\bf K},{\bf r}_{0}}$ and $a^\dag_{\pm{\bf K},{\bf r}_{0}}$.

By using the linear spin-wave theory, we evaluate the dynamical spin structure factor given by
\begin{align}
S^{\mu\nu}({\bf q},\omega)
=
\frac{
\epsilon
}{\pi}
{\sum_{\bf K}}'
\sum_{p}
\frac{
\langle {\rm vac} |
S^\mu_{-{\bf q}}
| {\bf K} p \rangle
\langle {\bf K} p|
S^\nu_{\bf q}
| {\rm vac} \rangle
}{
(\omega - \varepsilon_{{\bf K}p})^2
+
\epsilon^2
},\label{eq:Sqw}
\end{align}
where $|{\rm vac} \rangle $ is the vacuum of the quasiparticles $b$, 
$| {\bf K} p \rangle  = b^\dag_{{\bf K}p} |{\rm vac} \rangle$, 
and $\epsilon$ corresponds to the relaxation rate. 
In inelastic neutron scattering experiments, the transverse components to the incident wave number ${\bf q}$ are observed~\cite{Squires2012}. 
Thus, we study the transverse component of the  dynamical spin structure factor defined as
\begin{align}
S_{\perp} ({\bf q},\omega) = S^{
\mu_{1}\mu_{1}
}({\bf q},\omega) + S^{
\mu_{2}\mu_{2}
}({\bf q},\omega),
\label{eq:Sqw_perp}
\end{align}
where 
$\mu_{1}$ and 
$\mu_{2}$ are the two orthogonal directions perpendicular to ${\bf q}$,
e.g., $
\mu_{1}
=y$ and $
\mu_{2}
=z$ for ${\bf q} \parallel \hat{\bf x}$.
Furthermore, using a polarized neutron beam, 
two transverse components can be decomposed by measuring the spin-flip  and non-spin-flip cross sections.


\section{Results}\label{sec:results}

\subsection{One-dimensional magnetic helices}\label{sec:result:1d}
First, we present the results for the 1D HS states. 
In Sec.~\ref{sec:analytic1d}, we discuss the case of the isotropic symmetric interaction, where the stable state has a uniform twist. 
In this case, we can derive the analytic forms of the spin-wave dispersion 
and the dynamical spin structure factor. 
We discuss their dependences on the interaction range $\gamma$, including the limit of $D\to0$.
In Sec.~\ref{sec:anisotropy1d}, we numerically show that the anisotropy $\Delta$ makes the twist of the HS state nonuniform, accordingly, modulates the excitation spectra. 
Finally, in Sec.~\ref{sec:mode}, we study the lowest-energy excitation mode.

\subsubsection{Uniform helical spin state in the isotropic case}\label{sec:analytic1d}
\begin{figure}[H]
  \centering
  \includegraphics[trim=0 80 0 0, clip,width=\columnwidth]{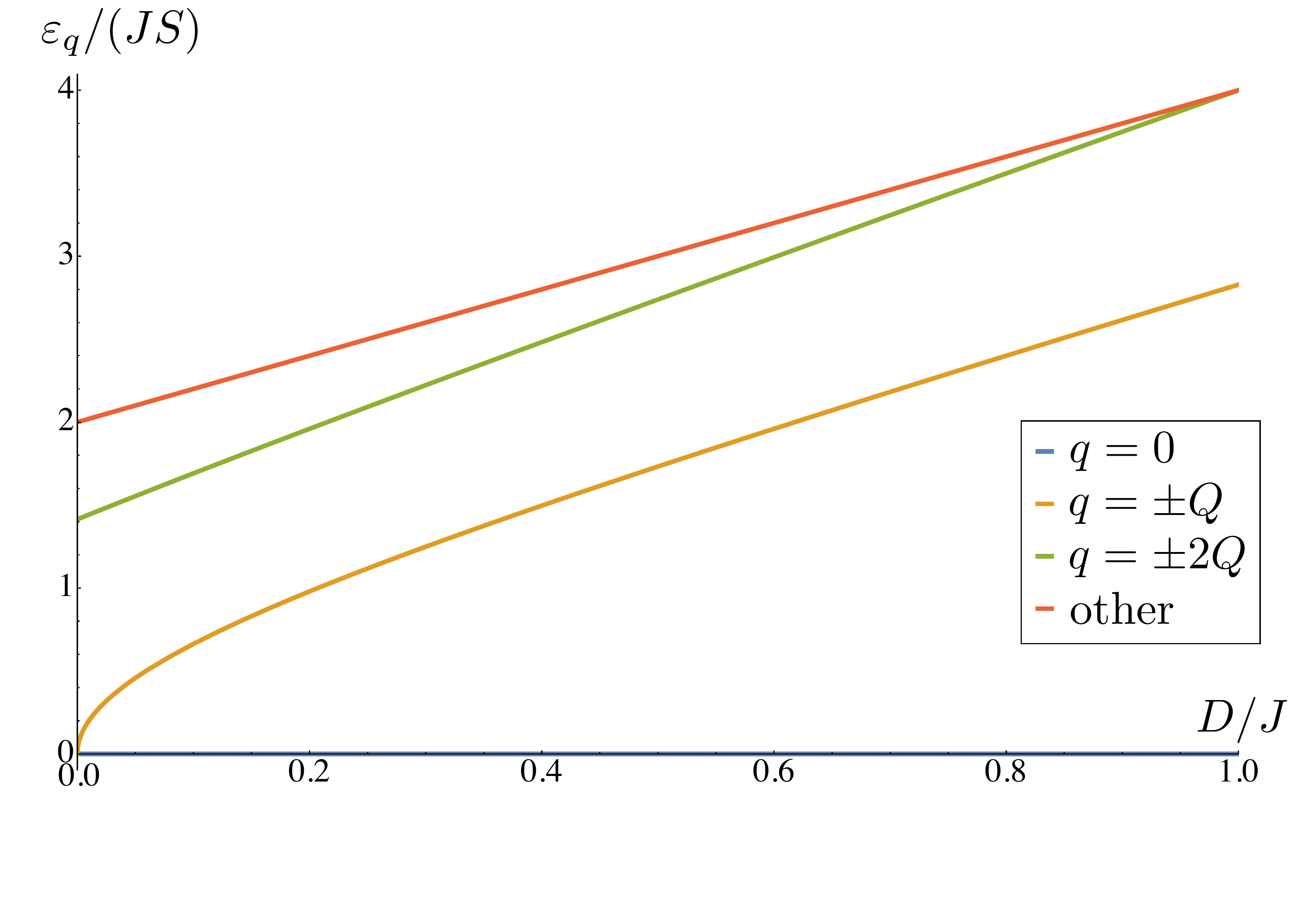}
  \caption{
	$D$ dependence of the spin excitation energy $\varepsilon_q$ of the infinite-range model ($\gamma=0$) with the isotropic symmetric interactions ($\Delta=0$)
	in one dimension for
	$q=0$ (blue),
	$q=\pm  Q $ (orange),
	$q=\pm 2  Q$ (green), and the other generic $q$ 
	(red) [Eq.~\eqref{eq:eps1d}].
  }
  \label{fig08}
\end{figure}
\begin{figure}[H]
  \centering
  \includegraphics[trim=0 0 0 0, clip,width=\columnwidth]{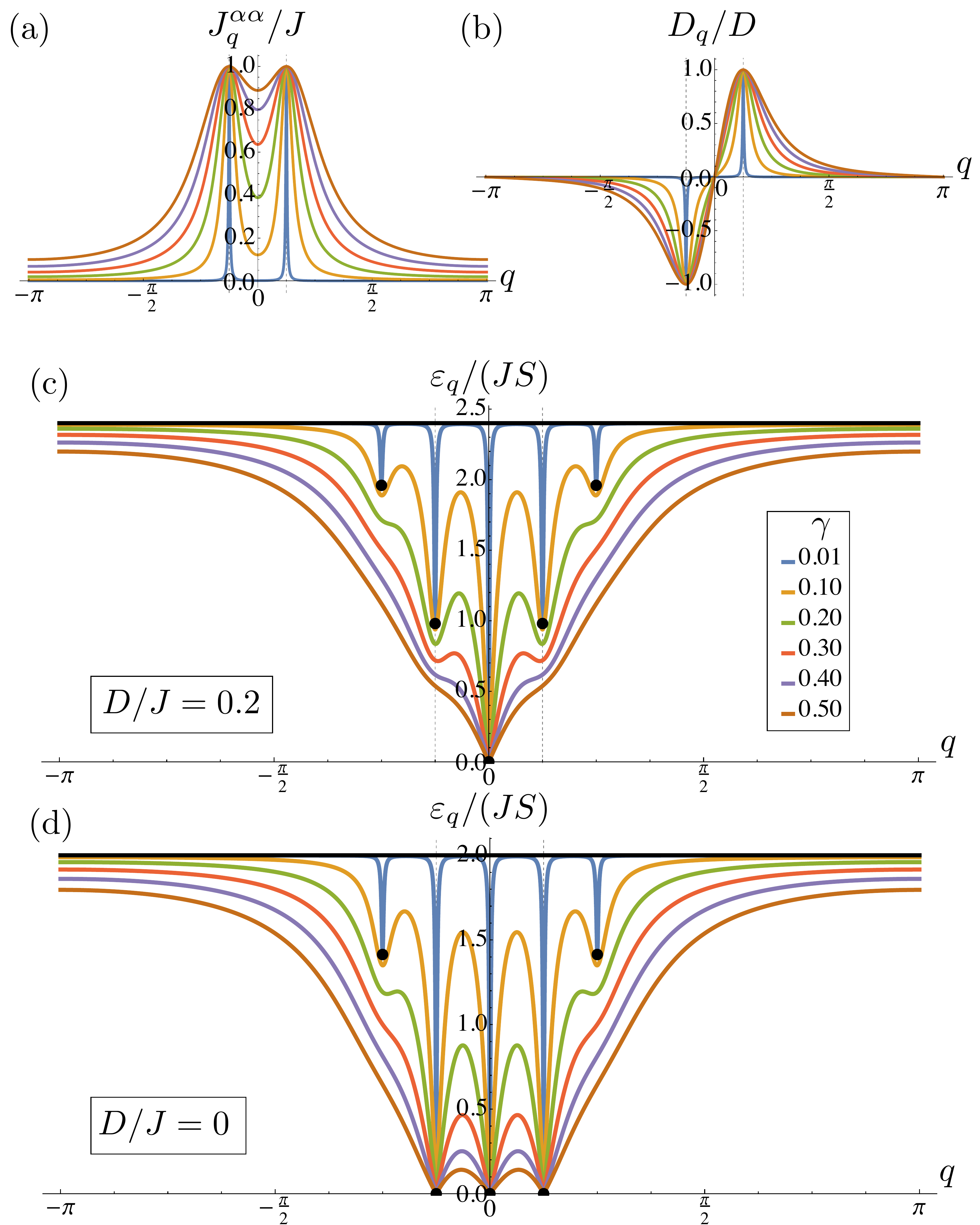}
  \caption{
    Interaction range $\gamma$ dependences of (a) the symmetric interaction $J_q=J^{\alpha\alpha}_q$, 
    (b) the antisymmetric interaction $D_q$, 
    and [(c) and (d)] the spin-wave dispersion $\varepsilon_q$ for the isotropic case ($\Delta=0$) with $\Lambda = 16$ in one dimension.
    The antisymmetric interaction is set to (c) $D/J=0.2$ and (d) $D/J=0$.
    The black dots and lines in (c) and (d) represent the results in the infinite-range limit of $\gamma \to 0$; see also Fig.~\ref{fig08}.  
  }
  \label{fig09}
\end{figure}
\begin{figure*}[ht]
  \centering
  \includegraphics[trim=0 0 0 0, clip,width=\textwidth]{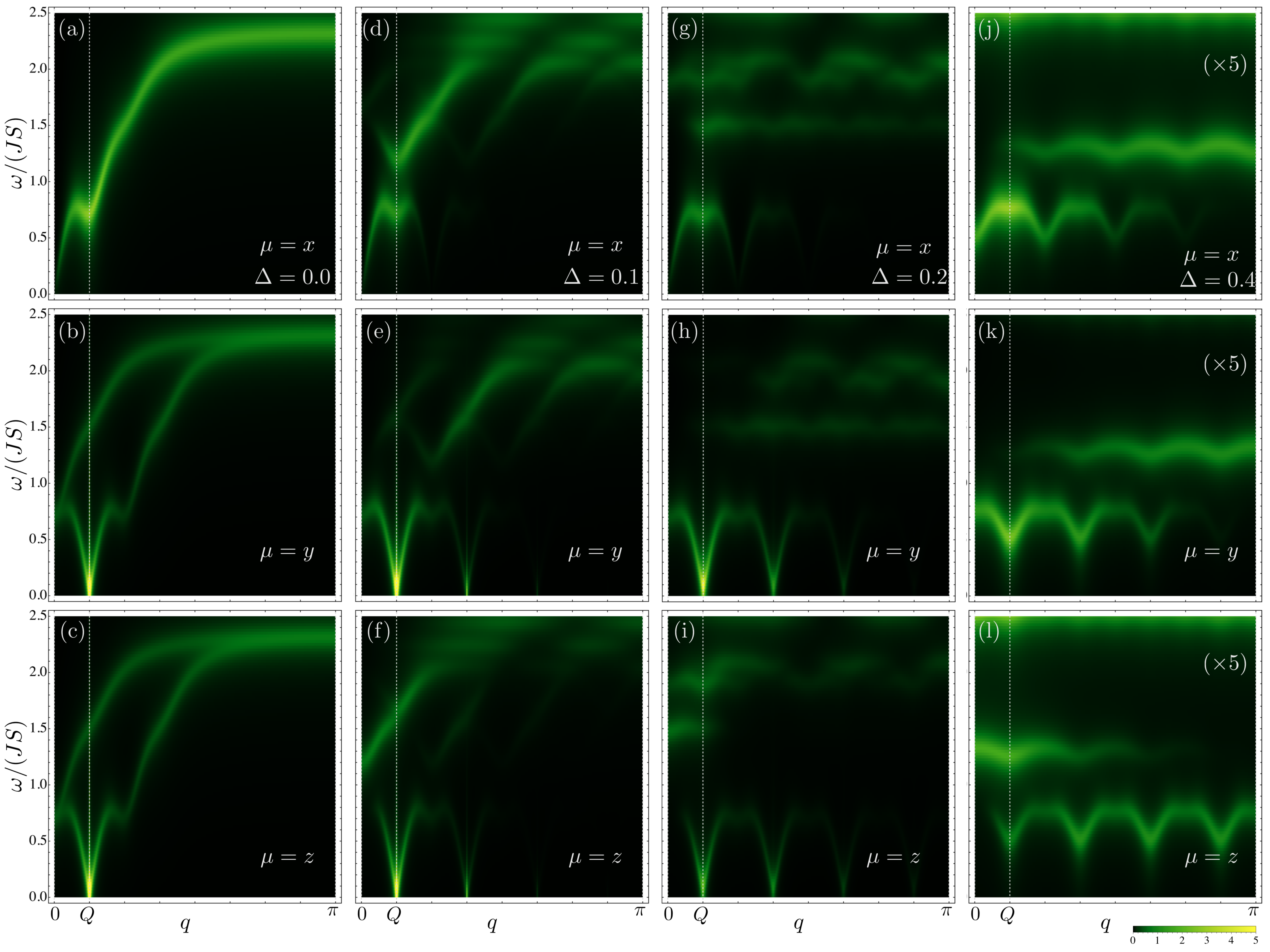}
  \caption{
  Anisotropy dependence of the dynamical spin structure factor
$S^{\mu\mu}(q,\omega)/S$ for the proper-screw HS state in one dimension.
  The anisotropy $\Delta$ is set to [(a)--(c)] $\Delta = 0$ (the isotropic case),
  [(d)--(f)] $\Delta = 0.1$, [(g)--(i)] $\Delta = 0.2$, and [(j)--(l)] $\Delta = 0.4$.
  (a,d,g,j), (b,e,h,k), and (c,f,i,l) show the $\mu = x$, $y$, and $z$ components, respectively. 
  The color bar is common to all plots whereas the data for $\Delta = 0.4$ in (j--l) are plotted by multiplying a factor of 5 for better visibility.
  The parameters are taken as $\Lambda = 16$, $D/J=0.2$, $\gamma=0.3$, and $\epsilon = 0.1$.
  }
  \label{fig10}
\end{figure*}
We begin with a HS state in one dimension which is stable when the symmetric interactions are isotropic, namely, $\Delta=0$ in Eq.~\eqref{eq:anisotropy}.
We here consider a proper-screw spin state given by
\begin{align}
{\bf S}_{\ell} =\left[0,\sin ( Q  \ell),\cos ( Q \ell) \right]^{\rm t},
\label{eq:PSspinstate}
\end{align}
and derive the analytic forms of the dispersion of spin excitation and the dynamical spin structure factor.

The Hamiltonian in Eq.~\eqref{eq:H} reads 
\begin{align}
\mathcal{H}_{{\bf q}} =&
-\left[
 J_{q}
{\bf S}_{q} \cdot {\bf S}_{-q}
+
 i D_{q}
\left(
{S}^y_{q} {S}^z_{-q} -{S}^z_{q} {S}^y_{-q} 
\right)
\right]
\nonumber\\
=&
- \frac{1}{L}
\sum_{\ell, \ell'}
\left[
 J_{q}
{\bf S}_{\ell} \cdot {\bf S}_{\ell'}\right.\notag\\
&\qquad\qquad +\left.
 i D_{q}
\left(
{S}^y_{\ell} {S}^z_{\ell'} - {S}^z_{\ell}{S}^y_{\ell'} 
\right)
\right]
e^{-iq(\ell -\ell')},
\label{eq:Hq1d}
\end{align} 
with $J_q\equiv J^{\alpha\alpha}_q$ and $D_q$ in Eqs.~\eqref{eq:frJq1d} and \eqref{eq:frDq1d}, respectively.
By substituting Eq.~\eqref{eq:PSspinstate} with the rotation of the local spin axes, namely, 
\begin{align}
{\bf S}_{\ell} =
\left(
\begin{array}{ccc}
 0 & 1 & 0 \\
 -\cos ( Q  \ell) & 0 & \sin ( Q  \ell) \\
 \sin ( Q  \ell) & 0 & \cos ( Q  \ell) \\
\end{array}
\right)
\tilde{\bf S}_{\ell},
\label{eq:Srot}
\end{align}
and applying the Holstein-Primakoff transformation in Eq.~\eqref{eq:HPtrans},
we obtain the linear spin-wave Hamiltonian as
\begin{align}
\mathcal{H}_{\rm SW}=&
S\sum_{q\in{\rm 1BZ}} \!\! \left\{
\frac{J_q}{2} 
(a^{\;}_{-q} \!- a^\dag_{q})
(a^{\;}_{q} \!- a^\dag_{-q})
\!+\!
2(J+D)
a^\dag_{q}
a^{\;}_{q} \right. \nonumber\\
&-
\frac{J_q + D_q}{4}
[a^{\;}_{-(q -  { Q })} +  a^\dag_{q -  { Q }}]
[a^{\;}_{q -  { Q }} + a^\dag_{-(q -  { Q })}] \nonumber\\
&\left. -
\frac{J_q - D_q}{4}
[a^{\;}_{-( q + { Q })} + a^\dag_{ q + { Q }}]
[a^{\;}_{ q + { Q }} + a^\dag_{-( q + { Q })}] \right\}
\nonumber\\
=& 
{\sum_q}'
\left[
a^\dag_q~
a^{\;}_{-q}
\right]
\mathsf{A}_q
\begin{bmatrix}
a^{\;}_q\\
a^{\dag}_{-q}
\end{bmatrix}
+{\rm const}.,
\label{eq:HSW1d}
\end{align}
where
\begin{align}
a^\dag_{q} =& \frac{1}{\sqrt{L}}\sum_{\ell}
a^\dag_{\ell} e^{-i q \ell},\\
\mathsf{A}_q =&S
\left[
J_q 
\begin{bmatrix}
-1 &1 \\
1 &-1 \\
\end{bmatrix}
+2(J+D)
\begin{bmatrix}
1 &0 \\
0 &1 \\
\end{bmatrix}
\right.\nonumber\\
&\left.
-\frac{J_{q+ Q } + D_{q+ Q }+J_{q- Q } - D_{q- Q }}{2}
\begin{bmatrix}
1 &1 \\
1 &1 \\
\end{bmatrix}
\right],
\end{align}
and the sum $\sum'_q$ in Eq.~\eqref{eq:HSW1d} runs over a half of the first Brillouin zone (e.g.,
$q \in [0,\pi]$)\footnote{
Strictly speaking, special treatment is required when $q = 0$ or $q=\pi$, but in reality, the same result is obtained by considering the limit of $q \to +0$ or $q \to \pi - 0$.
}.
Note that no term linear to the bosonic operators appears
as long as the HS state in Eq.~\eqref{eq:PSspinstate} is energetically stable.
Using the Bogoliubov transformation, the Hamiltonian in Eq.~\eqref{eq:HSW1d} is diagonalized as
\begin{align}
&\mathcal{U}_q \mathsf{A}_q \mathcal{U}_q=
\begin{bmatrix}
\varepsilon_q & 0\\
0 & \varepsilon_q
\end{bmatrix},\label{eq:Btrans1}
\end{align}
with
\begin{align}
\mathcal{U}_q=
\begin{bmatrix}
\cosh \xi_q & \sinh \xi_q  \\
\sinh \xi_q  & \cosh \xi_q 
\end{bmatrix},\label{eq:Btrans2}
\end{align}
where
\begin{align}
&\xi_q =\frac{1}{4} 
\left[ {\rm ln} B_q - {\rm ln} C_q  \right],\label{eq:Btrans3}\\
&B_q = J + D -J_q,\label{eq:Btrans4}\\
&C_q = 
 J + D -
\frac{
J_{q+ Q } + D_{q+ Q }
+J_{q- Q } \! - D_{q- Q }
}{2}.\label{eq:Btrans5}
\end{align}
The excitation spectrum is obtained as
\begin{align}
\varepsilon_q=
2S\sqrt{ B_q C_q}.
\end{align}

Let us first consider the infinite-range limit ($\gamma=0$) [Eqs.~\eqref{eq:Jq1d} and \eqref{eq:Dq1d0}].
In this limit, $\varepsilon_q$ becomes $q$ independent as $\varepsilon_q=2S(J+D)$, 
except for the $\delta$-functional changes at $q=0$, $\pm  Q $, and $\pm 2  Q $, namely,
\begin{align}
\varepsilon_q=
\begin{cases}
0,& (q=0)\\
2S \sqrt{ D (J + D ) } , & (q=\pm  Q ) \\
S \sqrt{ (J + D ) (J+3D)} ,& (q=\pm 2  Q ) \\
2S (J + D) ,& ({\rm other}~q)
\end{cases}.
\label{eq:eps1d}
\end{align}
The results are plotted as functions of $D/J$ in Fig.~\ref{fig08}.
We note that the flat dispersion with excitation energy $2S(J+D)$ originates from the
term 
$2S(J+D) \sum_\ell a^\dag_\ell a^{\;}_\ell$, 
indicating that the corresponding excitations are the local ones with reduction of 
the $\tilde{S}^z_\ell$ component at every site [see Eq.~\eqref{eq:HPtrans}].
This is a pathological feature of the infinite-range model.

Next, let us consider the finite-range model while changing the interaction range $\gamma$ [Eqs.~\eqref{eq:frJq1d} and \eqref{eq:frDq1d}].
Figures~\ref{fig09}(a) and \ref{fig09}(b) show $\gamma$ dependences of $J_q=J^{\alpha\alpha}_q$ and $D_q =|{\bf D}_q| $, respectively, for $\Lambda = 16$.
While increasing $\gamma$, the distributions of $J_q$ and $D_q$ in $q$ space get wider and 
qualitatively approach those of the model with the nearest-neighbor interactions only:
$J_q = J \cos q$ and $D_q = D \sin q$. 
Figure~\ref{fig09}(c) shows the excitation spectrum $\varepsilon_q$ at $D/J=0.2$. 
We find that the spikes at $q=0$, $\pm Q$, and $\pm 2Q$ for $\gamma=0$ are broadened by increasing $\gamma$; 
$\varepsilon_q$ is always zero at $q=0$, 
accompanied by a linear dispersion around the gapless point for nonzero $\gamma$. 

Meanwhile, as indicated in Fig.~\ref{fig08}, the spikes at $q=\pm Q$ for $\gamma=0$ also come down to zero energy when $D\to 0$.
Figure~\ref{fig09}(d) shows $\varepsilon_q$ in this limit. 
In this case, the broadening by nonzero $\gamma$ gives rise to gapless linear excitations at not only
$q=0$ but also $q=\pm  Q $.
These three gapless modes are commonly seen in the HS states appearing in spin models without the antisymmetric interactions, 
such as a $J_1$-$J_2$ model in one dimension~\cite{Cooper1962,Hasegawa_2010}.

Last, we derive the analytic form of the dynamical spin structure factor $S^{\mu\mu}(q,\omega)$ defined by Eq.~\eqref{eq:Sqw}. 
Within the linear spin-wave theory, by using Eqs.~\eqref{eq:HPtrans} and \eqref{eq:Srot}, we replace the spin operators $S^\mu_q$ by linear combinations of the bosonic operators as
\begin{align}
\begin{bmatrix}
S^x_q\\
S^y_q\\
S^z_q
\end{bmatrix}
\to
\sqrt{\frac{S}{2}}
\begin{bmatrix}
\frac{1}{i}
(a^{\;}_{-q}-
a^{\dag}_q)
\\
-\frac{1}{2} 
( a^{\;}_{-(q- Q )}
+a^\dag_{  q- Q  }
+a^{\;}_{-(q+ Q )}
+a^\dag_{  q+ Q  })
\\
\frac{1}{2i}
( a^{\;}_{-(q- Q )}
+a^\dag_{  q- Q  }
-a^{\;}_{-(q+ Q )}
-a^\dag_{  q+ Q  })
\end{bmatrix}.
\end{align}
Then, 
using the Bogoliubov transformation~[Eqs.~\eqref{eq:Btrans1}--\eqref{eq:Btrans5}],
we obtain the diagonal components of the dynamical spin structure factor as
\begin{align}
S^{xx}(q,\omega)=&
\frac{S \epsilon}{2\pi} \!\!
\left[
\frac{
e^{-2 \xi_q} 
}{
(\omega - \varepsilon_{q})^2
+
\epsilon^2
}
\right]
,\label{eq:Sxx:}
\\
S^{yy}(q,\omega)=& S^{zz}(q,\omega)\notag\\
=&
\frac{S\epsilon}{8\pi} \!\!
\left[
\frac{
e^{2 \xi_{q+ Q }} 
}{
(\omega - \varepsilon_{q+ Q })^2
+
\epsilon^2
}
+
\frac{
e^{2 \xi_{q- Q }} 
}{
(\omega - \varepsilon_{q- Q })^2
+
\epsilon^2
}
\right]\!\!,
\label{eq:SyySzz:}
\end{align}
where $\xi_q$ is given by Eq.~\eqref{eq:Btrans3}. 
Noting
\begin{align}
e^{-2\xi_q} &=
\sqrt{
\frac{
2( J + D ) - J_{q+ Q } - D_{q+ Q } -J_{q- Q } + D_{q- Q }
}{
2( J + D -J_q )
}
}\notag\\
&\xrightarrow[]{q\to 0}
|q|,
\end{align}
we can show the asymptotic behaviors:
\begin{align}
\label{eq:Sxx_asymptotic}
&S^{xx}(q,\omega) 
\xrightarrow[]{q\to 0 
} |q|,\\
\label{eq:Syy_asymptotic}
&S^{yy}(q,\omega)=S^{zz}(q,\omega) 
\xrightarrow[]{q\to \pm  Q } |q\mp  Q |^{-1}.
\end{align}

Figures~\ref{fig10}(a), \ref{fig10}(b), and \ref{fig10}(c) show 
$S^{xx}(q,\omega)$, $S^{yy}(q,\omega)$, and $S^{zz}(q,\omega)$, respectively, for the model with $\Lambda = 16$, $D/J = 0.2$, and $\gamma = 0.3$; 
we take $\epsilon= 0.1$ in Eqs.~\eqref{eq:Sxx:} and \eqref{eq:SyySzz:}. 
Note that $S^{xx}(q,\omega)$ vanishes as $q\to 0$, while
$S^{yy}(q,\omega)=S^{zz}(q,\omega)$ diverge as $q\to \pm Q$, as shown in Eqs.~\eqref{eq:Sxx_asymptotic} and \eqref{eq:Syy_asymptotic}.
In the inelastic neutron scattering experiments, 
only the transverse components to the wave number 
${\bf q}=q \hat{\bf x}$,  
namely, $S^{yy}(q,\omega)$ and $S^{zz}(q,\omega)$, can be observed, as mentioned below Eq.~\eqref{eq:Sqw}. 
This means that for the proper-screw HS state the divergent behaviors at $q\to \pm Q$ in $S^{yy}(q,\omega)$ and $S^{zz}(q,\omega)$ are observable, but the $q$-linear mode around $q=0$ with increasing intensity for larger $q$ in $S^{xx}(q,\omega)$ cannot be observed. 
Note that when we consider a cycloidal HS   
state, in which the spins are rotated by $\pi/2$ about the $z$ axis from the proper-screw one (Fig.~\ref{fig01}), $S^{xx}(q,\omega)$ and $S^{yy}(q,\omega)$ are interchanged, and hence, the $q$-linear mode with increasing intensity for larger $q$ is observed in the $S^{yy}(q,\omega)$ component. 
Thus, the neutron scattering spectra are sensitive to the direction of the helical plane. 
Similar behaviors were discussed for short-range models~\cite{dosSantos2018,Weber2018}.

\subsubsection{Effect of magnetic anisotropy}\label{sec:anisotropy1d}
\begin{figure}[htb]
  \centering
  \includegraphics[trim=0 0 0 0, clip,width=\columnwidth]{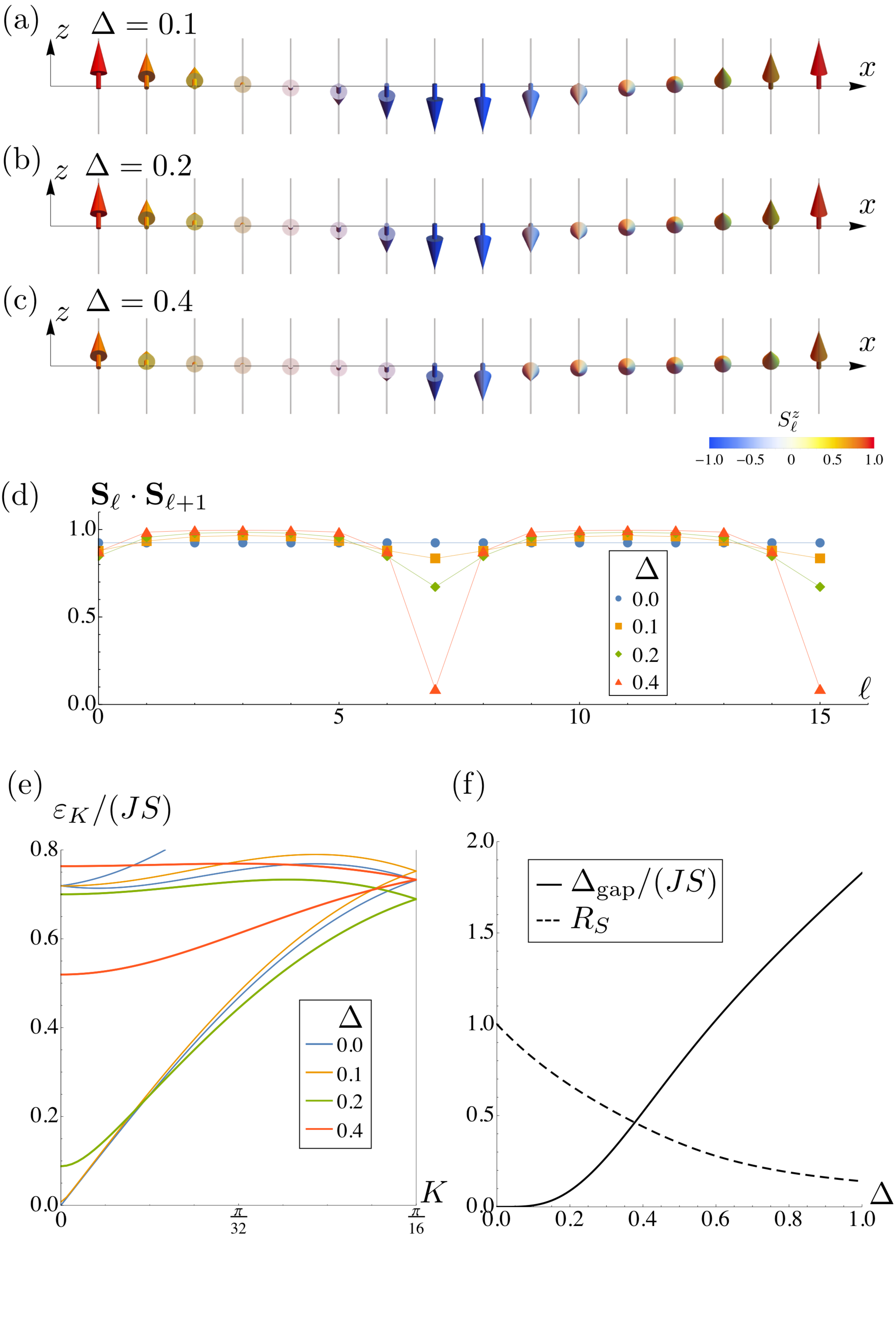}
  \caption{
  Effect of the anisotropy $\Delta$ in the symmetric interactions 
  [Eq.~\eqref{eq:anisotropy}] for the 1D finite-range model with $\Lambda = 16$, $D/J=0.2$, and $\gamma=0.3$.
  (a--c) Stable spin configurations for (a) $\Delta=0.1$, (b) $\Delta=0.2$, and  (c) $\Delta=0.4$.
  The color of arrows indicates the $z$ component of spin according to the color bar in (c).
  (d) Inner product of nearest-neighbor spins, ${\bf S}_\ell \cdot {\bf S}_{\ell+1}$, representing the spatial modulation of the spin twist.
  (e) Spin excitation spectra for different $\Delta$.
  (f) $\Delta$ dependences of the spin excitation gap $\Delta_{\rm gap}$ and the ratio of the Fourier components of spins, $R_S = |S^z_{ Q }|/|S^y_{ Q }|$.
  }
  \label{fig11}
\end{figure}
\begin{figure}[htb]
  \centering
  \includegraphics[trim=0 0 0 0, clip,width=\columnwidth]{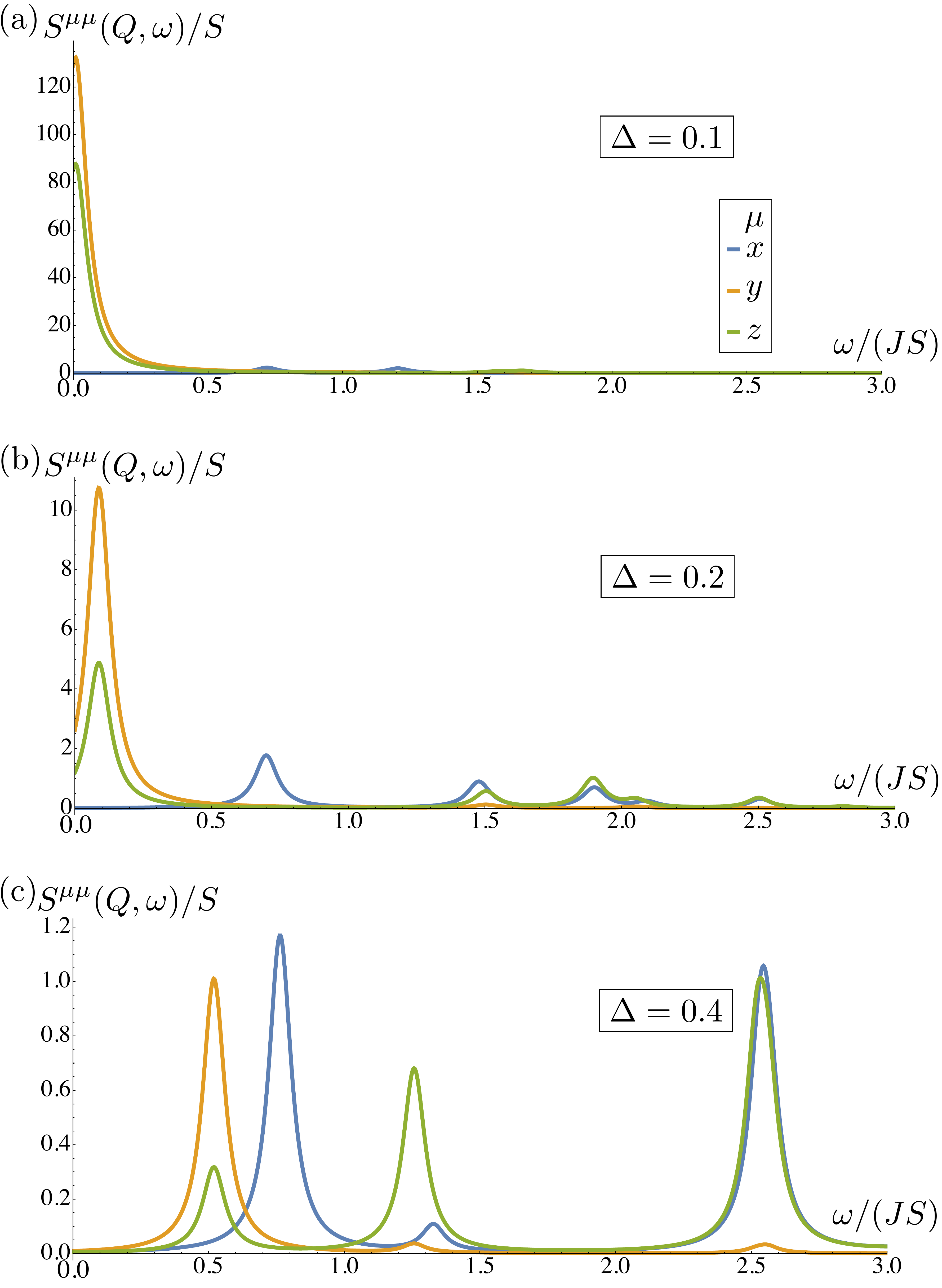}
  \caption{
  The dynamical spin structure factor $S^{\mu\mu}(q,\omega)/S$ at $q= Q $ for (a) $\Delta = 0.1$,
  (b) $\Delta = 0.2$, and (c) $\Delta = 0.4$.
  The model and parameters are the same as Fig.~\ref{fig11}.
  }
  \label{fig12}
\end{figure}
When we introduce the anisotropy $\Delta$ in the symmetric interactions as Eq.~\eqref{eq:anisotropy}, the proper-screw HS state is modulated from Eq.~\eqref{eq:PSspinstate}.
Figures~\ref{fig11}(a)--\ref{fig11}(c) show our numerical results for the stable spin configurations obtained by the variational calculation in Sec.~\ref{sec:variational}.
We find that the twist of the helix is modulated and becomes spatially nonuniform 
in the presence of the anisotropy $\Delta$, as more clearly shown in Fig.~\ref{fig11}(d).
This is because the spins tend to align to the $\pm \hat{\bf y}$ direction to gain energy for $\Delta>0$.
Indeed, we find that the ratio of the Fourier components of spins,
$R_S = |S^z_{ Q }|/|S^y_{ Q }|$, monotonically decreases while increasing $\Delta$, as shown in Fig.~\ref{fig11}(f).
We note that similar HS states with inhomogeneous twist were studied for a short-range model~\cite{Izyumov1983} and observed in CuB$_2$O$_4$~\cite{Roessli2001} and TbMnO$_3$~\cite{Kajimoto2004}.

Figure~\ref{fig11}(e) shows how the spin-wave dispersion is changed by $\Delta$.
When $\Delta > 0 $, the gap opens at $K=0$ ($K$ represents  
the wave number in the folded Brillouin zone as defined in Sec.~\ref{sec:LSWtheory}) and 
monotonically increases with increasing $\Delta$. 
The $\Delta$ dependence of $\Delta_{\rm gap}$ is plotted in Fig.~\ref{fig11}(f).

Figure~\ref{fig10} displays the dynamical spin structure factor for the modulated proper-screw HS states with several values of $\Delta$.
The spectra for $\Delta>0$ are gapped reflecting the spin-wave excitation, 
although the gap is small and hardly seen in the spectra for $\Delta=0.1$ and $0.2$. 
In addition, while increasing $\Delta$, the intensities at $q=\pm Q$ become weaker and the overall spectra become diffusive.
Looking more closely, we find that 
a nonzero $\Delta$ makes $S^{yy}(q,\omega)$ different from $S^{zz}(q,\omega)$;
$S^{yy}(q,\omega)$ becomes larger than $S^{zz} (q,\omega)$ in the low-energy part around $q=\pm Q$. 
This is more clearly seen in the $\omega$ dependence at $q=Q$ shown in Fig.~\ref{fig12}. 
The results are consistent with $R_S$ being smaller than 1 while increasing $\Delta$ [Fig.~\ref{fig11}(f)].
On the other hand, unlike $S^{yy}(Q,\omega)$ and $S^{zz}(Q,\omega)$, 
$S^{xx}(Q,\omega)$ takes the largest value at a higher energy around $\omega/(JS) \approx 0.7$ 
as shown in Figs.~\ref{fig10} and \ref{fig12}, 
whose energy scale roughly corresponds
to $\varepsilon_{\pm Q}$ in the infinite-range limit~[Eq.~\eqref{eq:eps1d}].
The transverse component of the dynamical spin structure factor, $S_{\perp}({\bf q},\omega)$ in Eq.~\eqref{eq:Sqw_perp},
for the proper-screw HS state is obtained by setting $\mu_1=y$ and $\mu_2=z$ as discussed in the end of Sec.~\ref{sec:analytic1d},
and then the results for the other HS states are obtained by using the corresponding rotations in spin space:
\begin{align}
&{\bf S}=\left(
S^x, S^y, S^z
\right), ~ \text{proper-screw HS state} \nonumber\\
&~~ \to
\begin{cases}
{\bf S}=\left(
-S^y, S^x, S^z
\right), &\text{cycloid(I) HS state}\\
{\bf S}=\left(
S^z, S^x, S^y
\right),
&\text{cycloid(II) HS state}
\end{cases}.
\end{align}
Since $S^{xx}(q,\omega)$ is (not) observed in the proper-screw (cycloidal) HS state, the presence or absence of the higher-energy intensity around $\omega \sim \varepsilon_{\pm Q}$
at ${\bf q}=Q\hat{\bf x}$ in the inelastic neutron scattering experiments 
can be an indicator for distinguishing the proper-screw and cycloidal HS states. 
In addition, it is worth noting that the spectra for $\Delta>0$ exhibit 
higher harmonics at $q=\pm 3  Q $, originating from the modulation of the spin configurations by the anisotropy.
Such satellite peaks were observed in neutron scattering experiments of CuB$_2$O$_4$~\cite{Roessli2001} and TbMnO$_3$~\cite{Kajimoto2004}.

\subsubsection{Mode analysis}\label{sec:mode}
\begin{figure}[ht]
  \centering
  \includegraphics[trim=0 0 0 0, clip,width=\columnwidth]{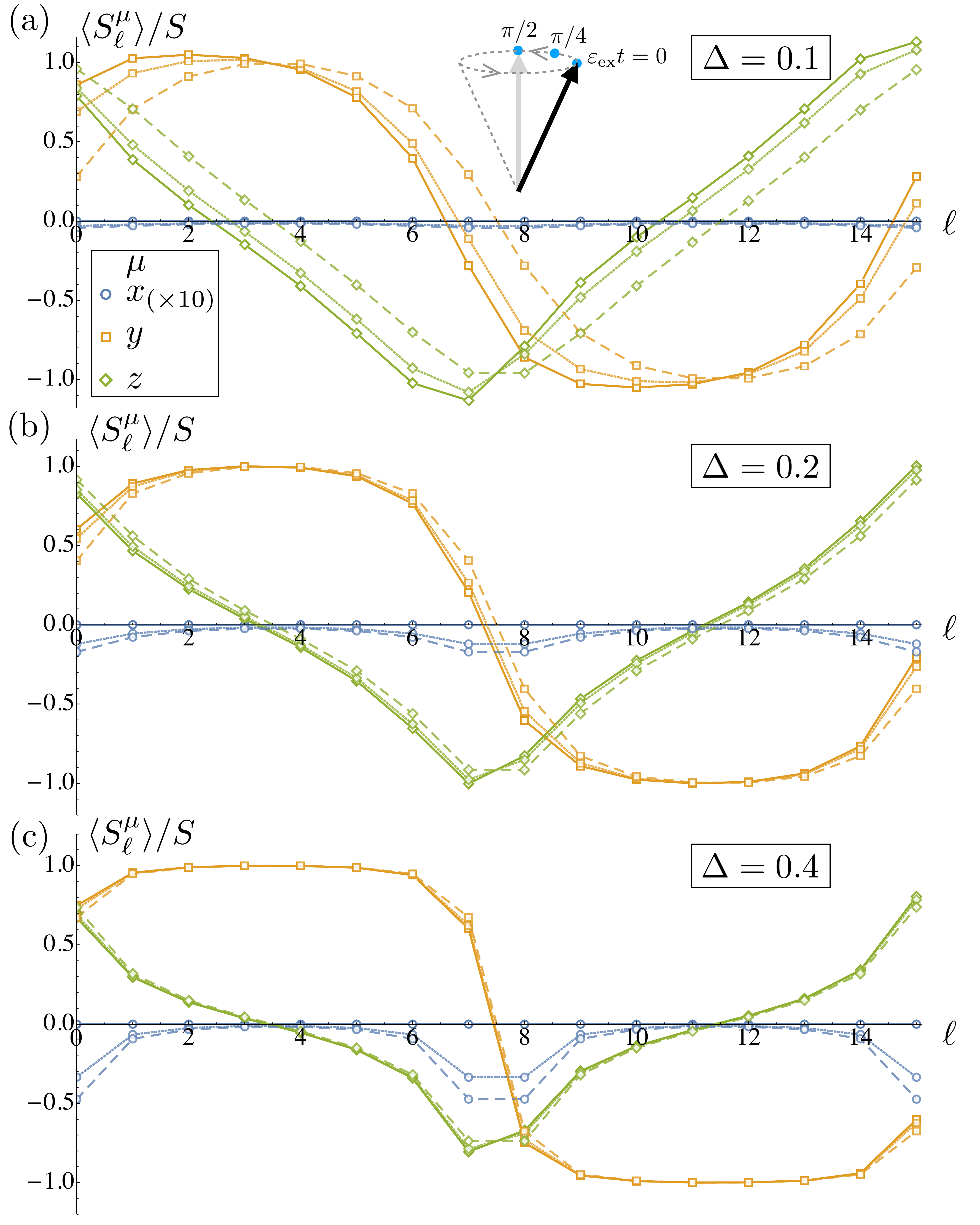}
  \caption{
  Time evolution of the lowest-energy excitation mode at $K=0$ for (a) $\Delta = 0.1$, (b) $\Delta = 0.2$, and (c) $\Delta = 0.4$.
  The data connected by the solid, dotted, and dashed lines indicate the spin configurations obtained by Eq.~\eqref{eq:deltaS} with $\varepsilon_{\rm ex} t = 0$, $\pi/4$, and $\pi/2$, respectively, for $\lambda=0.1$. 
  The model and parameters are the same as Fig.~\ref{fig11}.
  Inset of (a) shows a schematic view of a spin precession motion (black arrow) around the ground state (gray arrow). 
  }
  \label{fig13}
\end{figure}
Let us discuss the nature of the lowest-energy excitation mode.  
For this purpose, we consider a wave function 
\begin{align}
|\varphi \rangle = |{\rm vac} \rangle + 
\lambda e^{- i \varepsilon_{\rm ex} t}|{\rm ex} \rangle,
\label{eq:lowest-energy_wf}
\end{align}
 where $\lambda$ denotes a mixing between the ground state $|{\rm vac} \rangle$ (vacuum of magnons) and the lowest-energy excited state $|{\rm ex} \rangle$ at $K=0$:
$|{\rm ex} \rangle=|K=0,p=1\rangle$ with excitation energy of $\varepsilon_{\rm ex}\equiv \varepsilon_{K=0,p=1}$ [see Eq.~\eqref{eq:HSWdiag}]. 
For simplicity, in this section, we assume the large $S$ limit, in which $\langle {\rm vac} | \tilde{S}^z_\ell | {\rm vac} \rangle = S$.
Then, in the linear spin-wave theory, the expectation values of spins are computed as
\begin{align}
\langle \varphi | \tilde{S}^\mu_\ell |\varphi \rangle 
=&
\langle {\rm vac} | \tilde{S}^\mu_\ell | {\rm vac} \rangle\notag\\
&+
\lambda 
\left[
e^{-i \varepsilon_{{\rm ex}} t}
\langle {\rm vac} | \tilde{S}^\mu_\ell | {\rm ex} \rangle 
+
{\rm h.c.}
\right] 
+\mathcal{O}(
\lambda^2)\notag\\
\approx &
S\delta_{\mu z}
+
\frac{\lambda}{2} {\rm Re} \left[
e^{-i \varepsilon_{{\rm ex}} t}
\langle {\rm vac} | \tilde{S}^\mu_\ell | {\rm ex} \rangle 
\right]
=:\langle S_\ell^\mu \rangle.
\label{eq:deltaS}
\end{align}

Figure~\ref{fig13} shows the results of numerical calculations for $\Delta = 0.1$, $0.2$, and $0.4$ with $\lambda=0.1$.
At $t = 0$, all the spins for the excited state
$\varphi$ are in the helical plane, namely $\langle S_\ell^x\rangle = 0$. 
While increasing $t$, each spin shows an elliptically distorted precession, as schematically shown in the inset of Fig.~\ref{fig13}(a). 
After the precession by $\varepsilon_{\rm ex}t = \pi/2$, 
the $yz$ components of spins are indistinguishable from those of the ground state,
and only the $x$ component is different from the ground state.
These features are commonly seen regardless of $\Delta$.

The amplitude of the precession motion, however, strongly depends on $\Delta$.
When $\Delta$ is small, 
the spin components in the helical plane,
$\langle S_\ell^y \rangle$ and $\langle S_\ell^z \rangle$, 
show large motions, while the perpendicular component $\langle S_\ell^x \rangle$ changes much smaller, as exemplified in Fig.~\ref{fig13}(a); 
the excitation mode can be regarded as a phase shift of the helix. 
With an increase of $\Delta$, the changes of $\langle S_\ell^y \rangle$ and $\langle S_\ell^z \rangle$ ($\langle S_\ell^x \rangle$) are suppressed (enhanced), as shown in Fig.~\ref{fig13}(b).
For larger $\Delta$, the spins are almost pinned in the $\pm y$ directions and the amplitude of the precession becomes small, as shown in Fig.~\ref{fig13}(c); 
$\langle S_\ell^y \rangle$ and $\langle S_\ell^z \rangle$ show almost no change, 
while $\langle S_\ell^x \rangle$ oscillates near the regions where $\langle S_\ell^y \rangle$ changes its sign.

\subsection{Two-dimensional vortex crystals}\label{sec:result2d}
Next, we present the results for the 2D VCs.
In Sec.~\ref{sec:phasediagram2d}, we show the ground-state phase diagram for the 2D model in the limit of infinite-range interactions computed by the variational calculations.
In the phase diagram, we find that the anisotropy $\Delta$ favors the 2D VCs.
In Sec.~\ref{sec:vc2d}, we examine the details of the 2D VCs and their stabilization mechanism.
In Sec.~\ref{sec:Int_range_dep2d}, we discuss the dependence of the spin-wave dispersion on the interaction range $\gamma$.
Finally, in Sec.~\ref{sec:INS2d}, we discuss the dynamical spin structure factor computed by the linear spin-wave theory.

\subsubsection{Phase diagram}\label{sec:phasediagram2d}
\begin{figure}[H]
  \centering
  \includegraphics[trim=0 0 0 0, clip,width=\columnwidth]{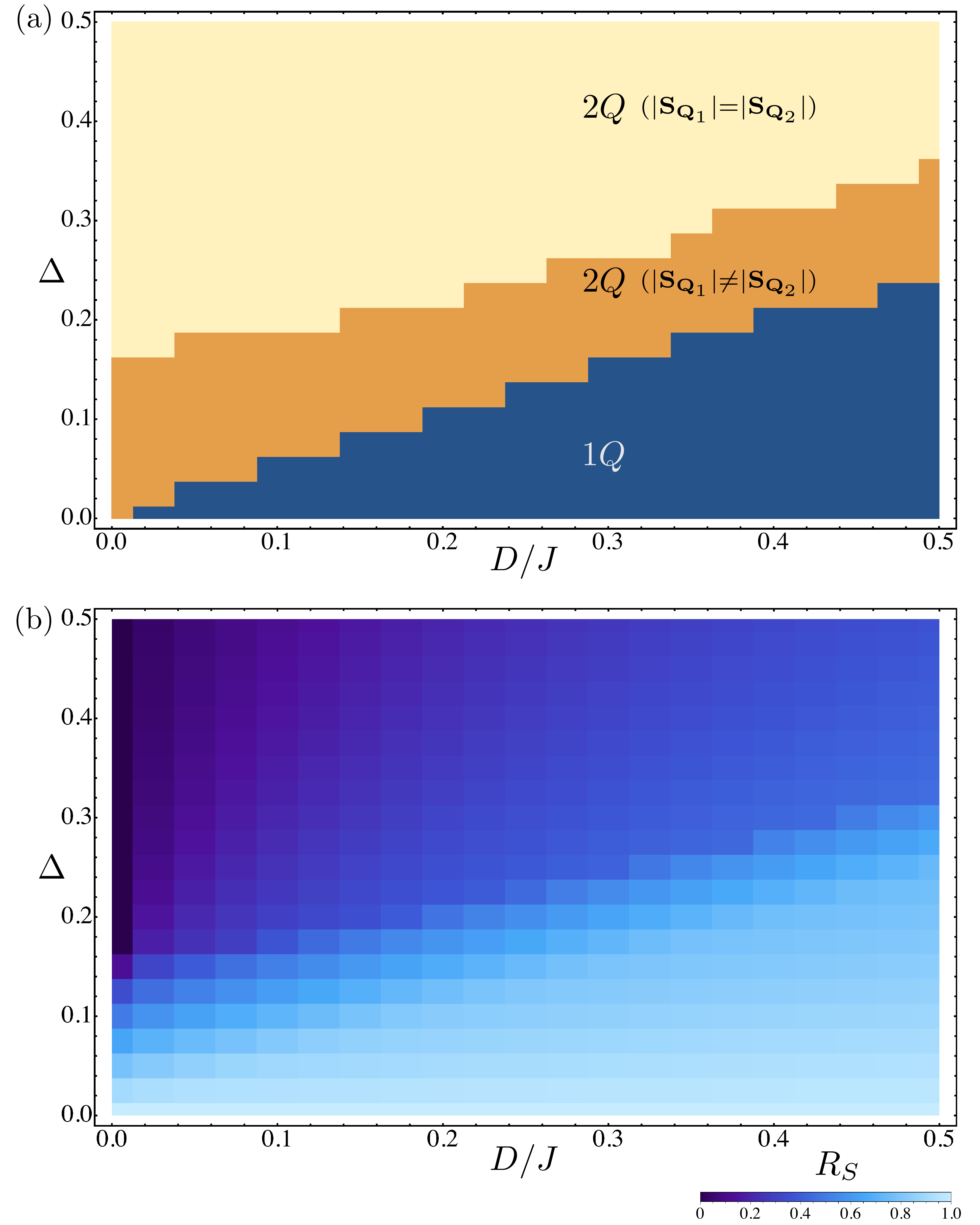}
  \caption{
  Variational results for the infinite-range model in two dimensions.
  We take $\Lambda=16$.
  (a) Phase diagram, including the $1Q$ state, the anisotropic $2Q$ state with $|{\bf S}_{{\bf Q}_1}| \neq |{\bf S}_{{\bf Q}_2}|$, and the isotropic $2Q$ state with $|{\bf S}_{{\bf Q}_1}| = |{\bf S}_{{\bf Q}_2}|$.
  (b) Contour plot of the ratio of the Fourier components of spins, $R_S \equiv |S^{\mu_{\perp,1}}_{{\bf Q}_{\rm max}}|/|S^{\mu_{\perp,2}}_{{\bf Q}_{\rm max}}|$, representing the ellipticity of the constituent spin helix; see the text for the definition.
  }
  \label{fig14}
\end{figure}
\begin{figure}[htb]
  \centering
  \includegraphics[trim=0 0 0 0, clip,width=\columnwidth]{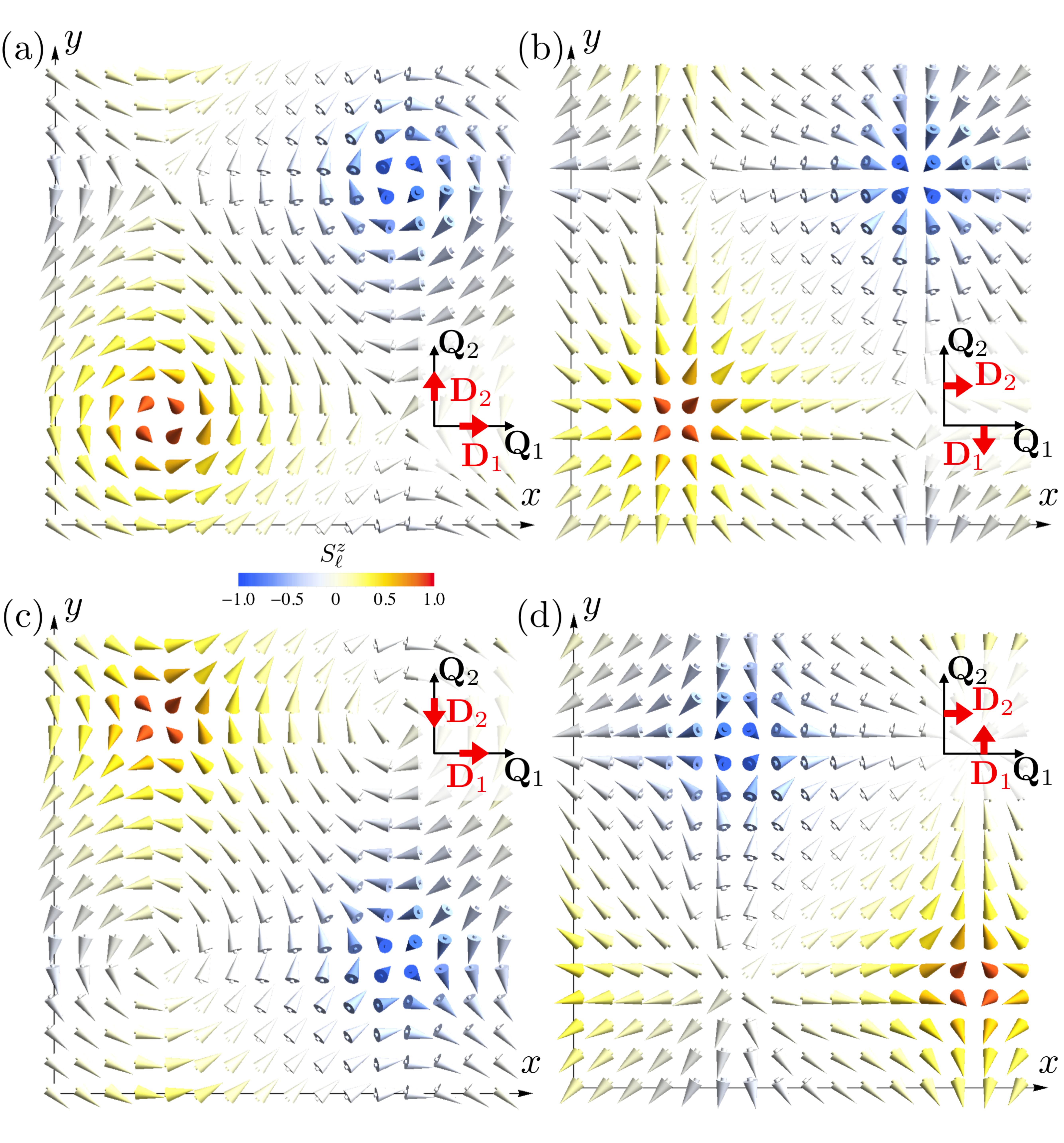}
  \caption{
  Isotropic $2Q$ spin states stabilized in the infinite-range model in two dimensions with $D/J=0.2$, $\Delta = 0.3$, and $\Lambda = 16$:
  (a) proper-screw(I), 
  (b) cycloid(I), 
  (c) proper-screw(II), and
  (d) cycloid(II) VCs. 
  The forms of $J^{\alpha\alpha}_{{\bf Q}_\eta}$ and ${\bf D}_{{\bf Q}_\eta}$ in each case are summarized in Table~\ref{tab1}. 
  The configurations of ${\bf D}_{{\bf Q}_\eta}$ as well as ${\bf Q}_\eta$ are shown in each figure.
  The color of arrows indicates the $z$ component of spin according to the color bar in (a).
  }
  \label{fig15}
\end{figure}
Figure~\ref{fig14}(a) shows the ground-state phase diagram for the infinite-range model in two dimensions~(Sec.~\ref{sec:inf_2D}) obtained by the variational calculations~(Sec.~\ref{sec:variational})
while changing $\Delta$ and $D/J$.
We take $\Lambda = 16$.
The result is common to all the settings of the proper-screw and cycloidal VCs listed in Table~\ref{tab1}.
We find three phases: the $1Q$ state, 
the anisotropic $2Q$ state where $|{\bf S}_{{\bf Q}_1}| \neq |{\bf S}_{{\bf Q}_2}|$, and
the isotropic $2Q$ state where $|{\bf S}_{{\bf Q}_1}| = |{\bf S}_{{\bf Q}_2}|$.
All the phase transitions among the three states look continuous.
In the isotropic case ($\Delta=0$), the system always stabilizes the $1Q$ state for $D>0$; 
it remains stable against nonzero $\Delta$, and the range becomes wider for larger $D$, as shown in Fig.~\ref{fig14}(a). 
In the larger $\Delta$ region, the system stabilizes the $2Q$ states, whose spin configurations are noncoplanar except for $D=0$. 
In the $2Q$ states, the constituent two spin helices are deformed from circular. 
To evaluate the ellipticity, we extend the ratio $R_S$ defined in Sec.~\ref{sec:anisotropy1d} to the present situation as $R_S=|S_{{\bf Q}_{\rm max}}^{\mu_{\perp,1}}|/|S_{{\bf Q}_{\rm max}}^{\mu_{\perp,2}}|$, where ${\bf Q}_{\rm max}$ is ${\bf Q}_\eta$ for larger $|{\bf S}_{{\bf Q}_\eta}|$, 
and $\mu_{\perp,1}$ and $\mu_{\perp,2}$ denote the $x$, $y$, or $z$  
directions perpendicular to ${\bf D}_{{\bf Q}_{\rm max}}$ 
satisfying $|S_{{\bf Q}_{\rm max}}^{\mu_{\perp,2}}|>|S_{{\bf Q}_{\rm max}}^{\mu_{\perp,1}}|$ 
(for instance,
$\mu_{\perp,1} =z$, and $\mu_{\perp,2} =y$ when ${\bf D}_{{\bf Q}_{\rm max}} \parallel \hat{\bf x}$ and $|S_{{\bf Q}_{\rm max}}^y|>|S_{{\bf Q}_{\rm max}}^z|$).
Note that $R_S=0$ in the isotropic $2Q$ state at $D=0$ and $\Delta \gtrsim 1.6$ because the spin state becomes coplanar ($S^z_{{\bf Q}_\eta} = 0$) composed of a superposition of two sinusoidal spin density waves with equal weight.
The calculated result of $R_S$ is plotted in Fig.~\ref{fig14}(b). 
We find that $R_S$ increases while increasing $D$ and decreasing $\Delta$.

\subsubsection{Vortex crystals}\label{sec:vc2d}

While the phase diagram is common to all the settings of the proper-screw and cycloidal VCs listed in Table~\ref{tab1}, 
the actual spin configuration in the $2Q$ state depends on the type of interactions. 
We present the variational results in Fig.~\ref{fig15}, focusing on the isotropic $2Q$ state at $D/J=0.2$ and $\Delta = 0.3$ ($R_S  \simeq 0.31$). 
Figure~\ref{fig15}(a) shows the stable spin configuration when taking $J^{\alpha\alpha}_{{\bf Q}_\eta}$ and ${\bf D}_{{\bf Q}_\eta}$ as Eqs.~\eqref{eq:JQeta2d1} and \eqref{eq:DQeta2d}, respectively. 
This is the proper-screw(I) VC. 
On the other hand, Fig.~\ref{fig15}(b) shows the spin configuration obtained for Eqs.~\eqref{eq:JQeta2d2} and \eqref{eq:DQeta2d2}, which is the cycloid(I) VC. 
Likewise, Figs.~\ref{fig15}(c) and \ref{fig15}(d) display the spin configurations for Eqs.~\eqref{eq:JQeta2d3} and \eqref{eq:DQeta2d3} and Eqs.~\eqref{eq:JQeta2d4} and \eqref{eq:DQeta2d4}, which are the proper-screw(II) and cycloid(II) VCs, respectively. 
Note that the VCs of type (I) [(II)] can be regarded as square lattices with a staggered arrangement of merons and antimerons with vorticity $+1$ ($-1$)~\cite{Gao2019}.
Similarly, the spin configurations of the anisotropic $2Q$ state also form VCs where the vortices are deformed (not shown). 

These VCs are stabilized by the anisotropy $\Delta$ in the symmetric interactions, as suggested in the phase diagram in Fig.~\ref{fig14}(a). 
This can be directly confirmed by calculating the spin components of ${\bf S}_{{\bf Q}_{\eta}}$. 
We find that all the VCs have large values of $|S_{{\bf Q}_1}^y|=|S_{{\bf Q}_2}^x|$ and $|S_{{\bf Q}_1}^x|=|S_{{\bf Q}_2}^y|$
for the proper-screw and cycloidal VCs, respectively: 
For example, at $D/J=0.2$ and $\Delta = 0.3$, we find $(|S_{{\bf Q}_1}^x|, |S_{{\bf Q}_1}^y|, |S_{{\bf Q}_1}^z|) \simeq (0, 7.4, 2.3)$ and
$(|S_{{\bf Q}_2}^x|, |S_{{\bf Q}_2}^y|, |S_{{\bf Q}_2}^z|) \simeq (7.4, 0, 2.3)$ for the proper-screw VCs, while 
$(|S_{{\bf Q}_1}^x|, |S_{{\bf Q}_1}^y|, |S_{{\bf Q}_1}^z|) \simeq (7.4, 0, 2.3)$ and
$(|S_{{\bf Q}_2}^x|, |S_{{\bf Q}_2}^y|, |S_{{\bf Q}_2}^z|) \simeq (0, 7.4, 2.3)$ for the cycloidal VCs.
This leads to the energy gain in the interaction terms, 
$-J(1+2\Delta)S_{{\bf Q}_1}^yS_{-{\bf Q}_1}^y$ and $-J(1+2\Delta)S_{{\bf Q}_2}^xS_{-{\bf Q}_2}^x$ in Eqs.~\eqref{eq:JQeta2d1} and \eqref{eq:JQeta2d3}, for the proper-screw VCs, 
while $-J(1+2\Delta)S_{{\bf Q}_1}^xS_{-{\bf Q}_1}^x$ and $-J(1+2\Delta)S_{{\bf Q}_2}^yS_{-{\bf Q}_2}^y$ in Eqs.~\eqref{eq:JQeta2d2} and \eqref{eq:JQeta2d4} for the cycloidal VCs.

\subsubsection{Interaction range dependence}\label{sec:Int_range_dep2d}
\begin{figure}[htb]
  \centering
  \includegraphics[trim=0 0 0 0, clip,width=\columnwidth]{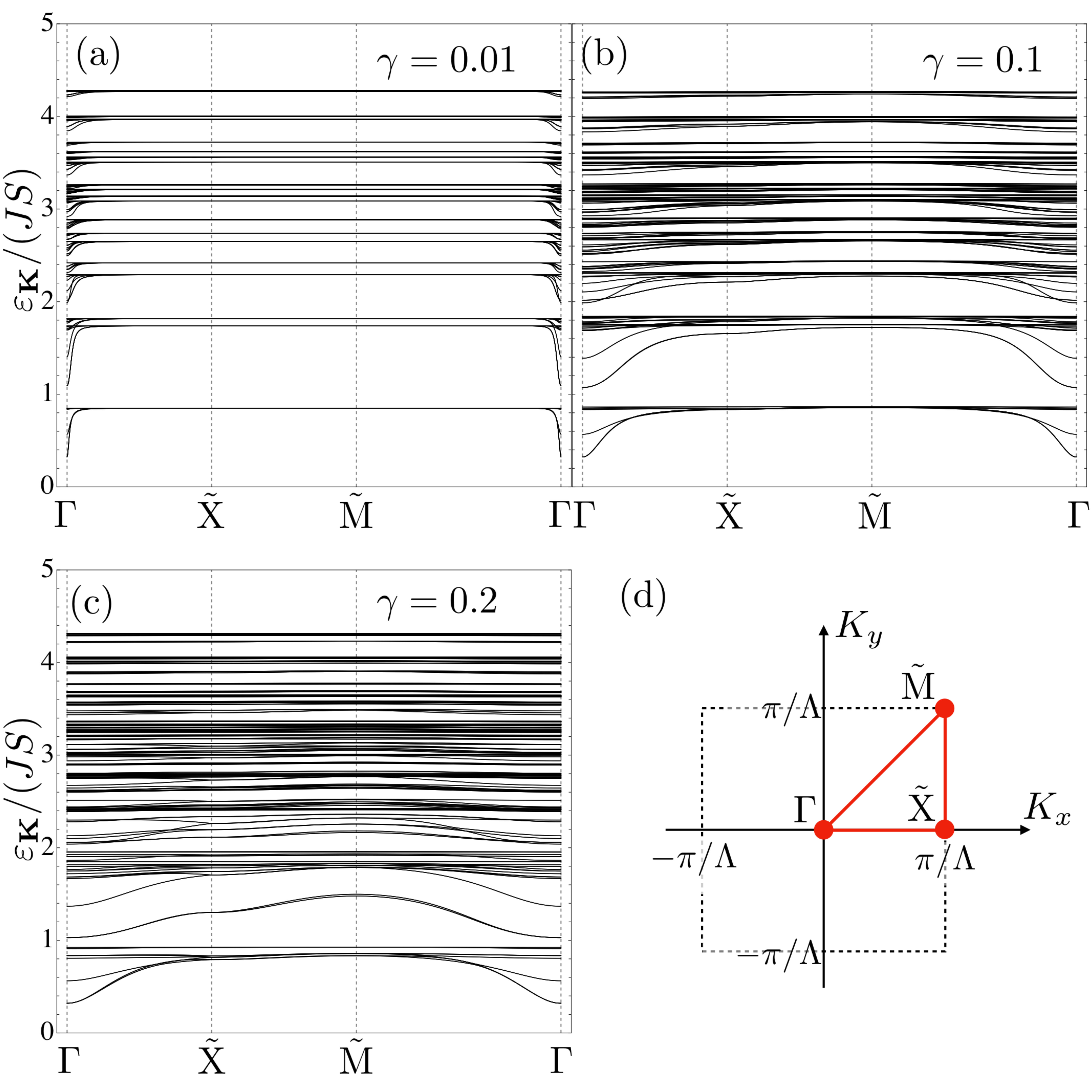}
  \caption{
    Spin excitation spectra in the isotropic $2Q$ spin states in two dimensions while changing the parameter for the interaction range:
    (a) $\gamma=0.01$, 
    (b) $\gamma=0.1$, and 
    (c) $\gamma=0.2$.
    We take $D/J=0.2$, $\Delta = 0.3$, and $\Lambda = 16$, as in Fig.~\ref{fig15}.
    (d) Symmetric lines in the folded Brillouin zone, used for the plots in (a--c). 
  }
  \label{fig16}
\end{figure}
Let us consider the finite-range model in two dimensions while changing the parameter for the interaction range, $\gamma$.
We perform the variational calculations for $\gamma>0$
starting from the stable spin configuration for the infinite-range limit of $\gamma=0$, 
which results in the solutions retaining the type of each VC with modulated spin configurations; 
while this procedure does not ensure that the resultant solution 
is the ground state, but it is, at least, a metastable state, for which we can compute the spin excitations by the linear spin-wave theory.
Figure~\ref{fig16} shows the spin-wave dispersion $\varepsilon_{\bf K}$ in the folded Brillouin zone for three different values of $\gamma$. 
The results are common to all the VCs in Fig.~\ref{fig15}.
When $\gamma$ is small enough, the excitation spectra are almost flat except around the $\Gamma$ point.
The multiple values of excitation energy arise from a nonuniform twist in VCs.
As in the 1D case~(Fig.~\ref{fig09}), while increasing $\gamma$, the dispersion becomes more dispersive, whereas the bandwidth is barely changed.

\subsubsection{Dynamical spin structure factor}\label{sec:INS2d}
\begin{figure}[h!]
  \centering
  \includegraphics[trim=0 50 0 0, clip,width=\columnwidth]{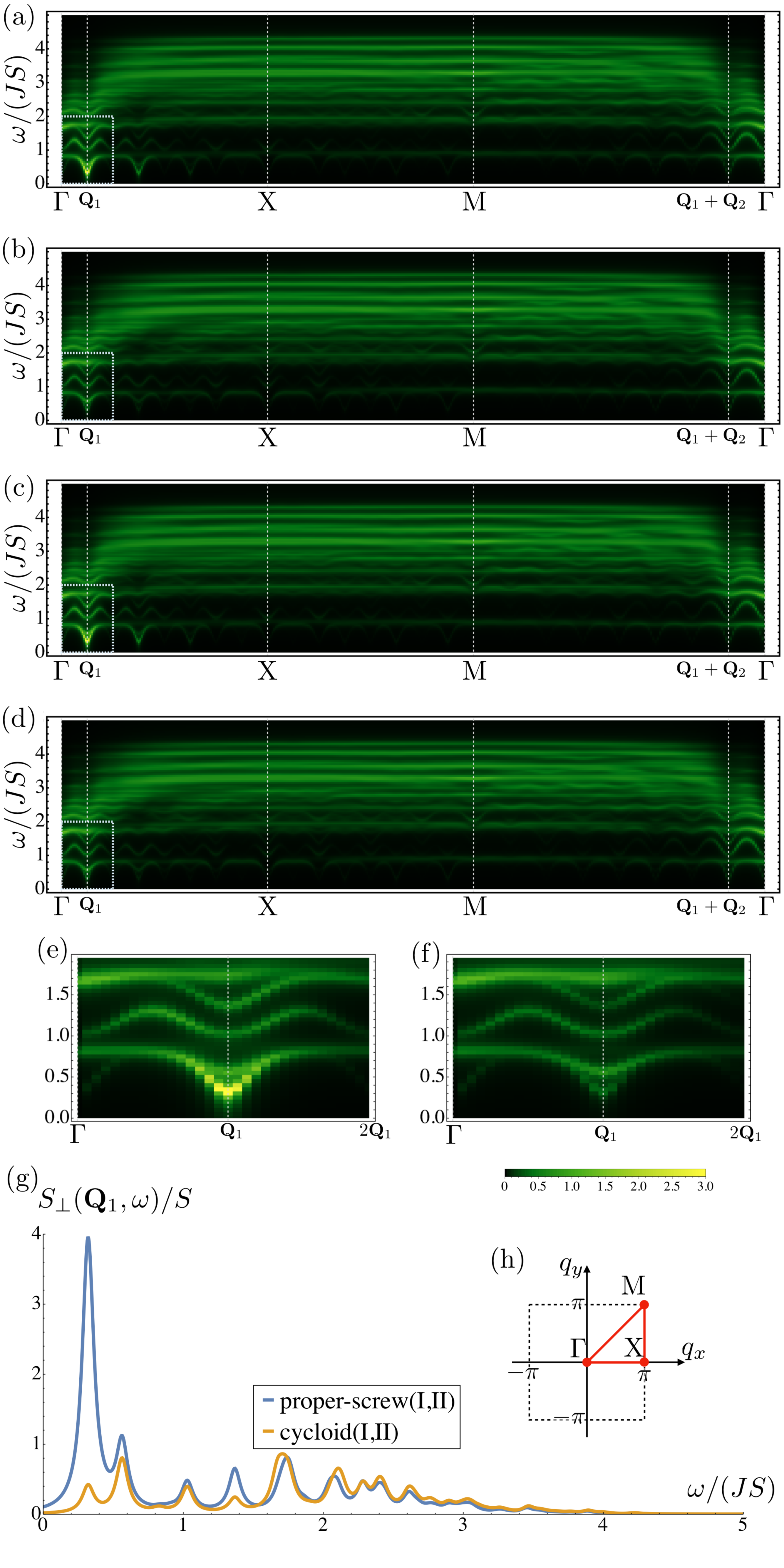}
  \caption{
  Transverse component of the dynamical spin structure factor, $S_{\perp}({\bf q},\omega)$ in Eq.~\eqref{eq:Sqw_perp}, for
 (a) proper-screw(I), (b) cycloid(I), (c) proper-screw(II), and (d) cycloid(II) VCs, plotted along the symmetric lines in the first Brillouin zone shown in (h). 
  (e) and (f) show the enlarged views of the dotted areas in (a,c) and (b,d), respectively.
  (g) $\omega$ dependence at ${\bf q}={\bf Q}_1$.
  The interaction range and the relaxation rate are taken as $\gamma=0.2$  and $\epsilon = 0.05$, respectively.
  The other parameters are $\Lambda = 16$, $D/J=0.2$, and $\Delta = 0.3$ as in Fig.~\ref{fig15}.
}
  \label{fig17}
\end{figure}

We discuss the transverse component of the dynamical spin structure factor, $S_{\perp}({\bf q},\omega)$ in Eq.~\eqref{eq:Sqw_perp},
for each VC, which is related to the observable in the inelastic neutron scattering experiment. 
Note that the direction of ${\bf q}$ is fixed along the $x$ direction in the $1Q$ case in Sec.~\ref{sec:result:1d}, where we discussed $S^{yy}({\bf q},\omega)$ and $S^{zz}({\bf q},\omega)$, but in the 2D case the ${\bf q}$ direction is rotated and the relevant spin components 
change with the direction. 
In the calculation, we first compute $\langle {\bf K}p|S^\mu_{{\bf q}} |{\rm vac} \rangle$ in Eq.~\eqref{eq:Sqw} using the model for the proper-screw(I) VC, 
and then obtain the results for the other VCs by using the corresponding rotations in spin space:
\begin{align}
&{\bf S}=
\left(
S^x, S^y, S^z
\right), ~ \text{proper-screw(I) VC} \nonumber\\
&~~\to
\begin{cases}
{\bf S}= 
\left( S^y, -S^x, S^z \right), &\text{cycloid(I) VC}\\
{\bf S}= 
\left( S^x, -S^y, -S^z \right),
&\text{proper-screw(II) VC}\\
{\bf S}= \left( S^y , S^x , -S^z \right),
&\text{cycloid(II) VC}
\end{cases}.
\end{align}

Figure~\ref{fig17} shows $S_{\perp}({\bf q},\omega)$ for the four types of VCs obtained at $D/J=0.2$, $\Delta = 0.3$, $\Lambda = 16$, and $\gamma=0.2$.
Here, we take $\epsilon=0.05$ in Eq.~\eqref{eq:Sqw}. 
The overall spectra look similar among the different VCs. 
In particular, as expected from the above calculation scheme, 
the spectra along the $\Gamma$--${\rm X}$ line are common to the two types of the proper-screw VCs [Figs.~\ref{fig17}(a) and \ref{fig17}(c)]; 
this holds also for the cycloid(I) and (II) VCs [Figs.~\ref{fig17}(b) and \ref{fig17}(d)]. 
In the same way, the spectra  
along the ${\rm M}$--$\Gamma$ line are common to the proper-screw(I) and cycloid(II) VCs [Figs.~\ref{fig17}(a) and \ref{fig17}(d)], and to the cycloid(I) and proper-screw(II) VCs [Figs.~\ref{fig17}(b) and \ref{fig17}(c)]. 
In addition, we note that the spectra along the ${\rm X}$--${\rm M}$ line are common to the two types of VCs for both proper-screw and cycloidal cases
(see Appendix~\ref{sec:appA}).

On the other hand, a stark difference between the proper-screw and cycloidal VCs is found along the $\Gamma$--${\rm X}$ line, especially in the vicinity of ${\bf q}={\bf Q}_1$:
The intensity of the lowest-energy excitation mode is much larger for the proper-screw VCs [Fig.~\ref{fig17}(e)] than the cycloidal VCs~[Fig.~\ref{fig17}(f)].
This is more clearly shown in the $\omega$ dependence at ${\bf q}={\bf Q}_1$ in Fig.~\ref{fig17}(g).
The large difference is 
consistent with the small $R_S$ plotted in Fig.~\ref{fig14}(b), since the ratio between the intensities at ${\bf q}={\bf Q}_1$ and $\omega=\varepsilon_{{\bf Q}_1\;p=1}$ is well approximated by $R_S^2$.
The reason is as follows.
The intensity is computed from the dynamical spin structure factor as $S_\perp({\bf Q}_1,\omega)=S^{yy}({\bf Q}_1,\omega)+S^{zz}({\bf Q}_1,\omega)$ for the two types of the proper-screw VCs, 
while $S_\perp({\bf Q}_1,\omega)=S^{xx}({\bf Q}_1,\omega)+S^{zz}({\bf Q}_1,\omega)$ for those of the cycloid.
At $\omega = \varepsilon_{{\bf Q}_1,p=1}$, $S_\perp({\bf Q}_1,\omega)$ is dominated by $S^{yy}({\bf Q}_1,\omega)$ [$S^{zz}({\bf Q}_1,\omega)$] for the proper-screw (cycloid) VCs.
Since the frequency integral of the dynamical spin structure factor corresponds to the static spin structure factor, i.e., $\int S^{\mu\mu}({\bf q},\omega) d\omega \propto |S^{\mu}_{\bf q}|^2$,
the intensity ratio is approximately given by $|S^{z}_{{\bf Q}_1}|^2/|S^{y}_{{\bf Q}_1}|^2=R_S^2$. 
For the present parameter set, $R_S \simeq 0.31$ as shown in Sec.~\ref{sec:vc2d}, leading to $R_S^2 \approx 0.10$. 
The value well explains the peak difference in Fig.~\ref{fig17}(g).
Thus, such a difference around ${\bf q}={\bf Q}_1$ could be useful to 
distinguish the proper-screw and cycloid types of VCs in experiments.

We note that it is rather difficult to distinguish the type (I) and (II) from the spectra, in both proper-screw and cycloidal cases. 
There is, however, a noticeable difference near $({\bf q},\omega) \approx ( {\bf Q}_1+{\bf Q}_2, 1.8 JS )$ along the ${\rm M}$--$\Gamma$ line, as shown in Figs.~\ref{fig17}(a)--\ref{fig17}(d).

\subsection{Three-dimensional hedgehog lattices}\label{sec:result3d}
Finally, we present the results for the 3D HLs.
The structure of the following sections is similar to that in Sec.~\ref{sec:result2d} for the 2D case: 
the variational phase diagram in Sec.~\ref{sec:phasediagram3d}, 
the details of the 3D HLs and their stabilization mechanism in Sec.~\ref{sec:hl3d}, 
the spin-wave dispersion while changing the interaction range in Sec.~\ref{sec:Int_range_dep3d}, and 
the dynamical spin structure factor in Sec.~\ref{sec:INS3d}.

\subsubsection{Phase diagram}\label{sec:phasediagram3d}
\begin{figure}[ht]
  \centering
  \includegraphics[trim=0 0 0 0, clip,width=\columnwidth]{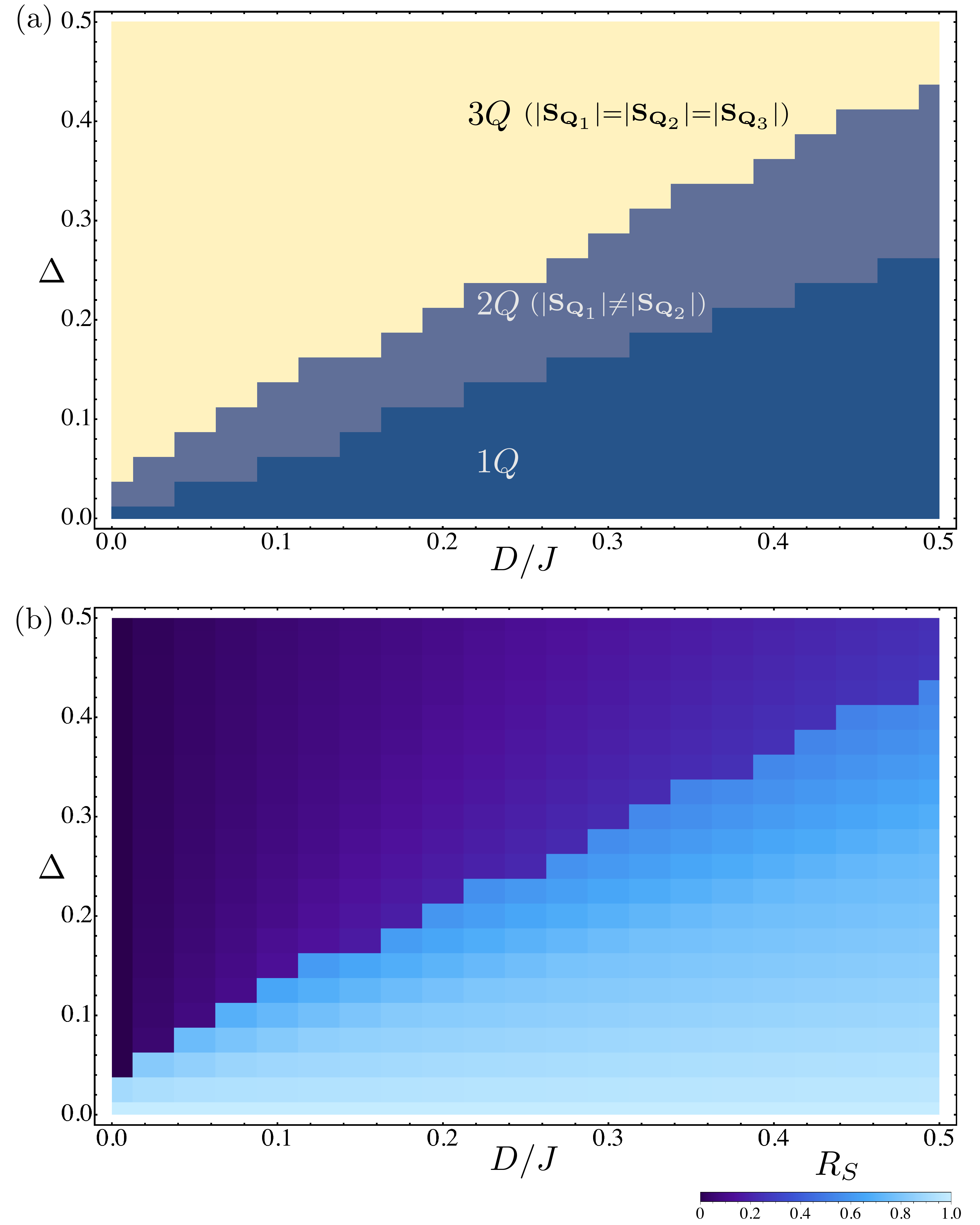}
  \caption{
  Variational results for the infinite-range model in three dimensions.
  We take $\Lambda=12$. 
  (a) Phase diagram, which includes the $1Q$ state, the anisotropic $2Q$ state with $|{\bf S}_{{\bf Q}_1}| >  |{\bf S}_{{\bf Q}_2}|\neq 0$ 
  and ${\bf S}_{{\bf Q}_3} = 0$ (the cyclic permutations of ${\bf Q}_\eta$ are energetically degenerate)
  and the isotropic $3Q$ state with $|{\bf S}_{{\bf Q}_1}| = |{\bf S}_{{\bf Q}_2}|= |{\bf S}_{{\bf Q}_3}|$. 
  (b) Contour plot of the ratio of the Fourier components of spins, $R_S$, defined in Sec.~\ref{sec:phasediagram2d} and Fig.~\ref{fig14}.
  }
  \label{fig18}
\end{figure}
Figure~\ref{fig18}(a) shows the ground-state phase diagram for the infinite-range model in three dimensions (Sec.~\ref{sec:inf_3D}) obtained by the variational calculations~(Sec.~\ref{sec:variational}) while changing $\Delta$ and $D/J$.
We here take $\Lambda=12$. 
The phase diagram is common to all the settings of the proper-screw and cycloid HLs listed in Table~\ref{tab1}.
We find three phases: the $1Q$ state, 
the anisotropic $2Q$ state where one of three $|{\bf S}_{{\bf Q}_\eta} |$ is zero 
and the other two have nonzero different values,
and
the isotropic $3Q$ state where $|{\bf S}_{{\bf Q}_1}| = |{\bf S}_{{\bf Q}_2}|= |{\bf S}_{{\bf Q}_3}|$.
The phase transition between the $3Q$ and $2Q$ states is discontinuous,
while that between $2Q$ and $1Q$ looks continuous. 
Similar to the 2D case, the system stabilizes the $1Q$ state for $D>0$ and small $\Delta$, as shown in Fig.~\ref{fig18}(a). 
In the larger $\Delta$ region, the system stabilizes the $2Q$ and $3Q$ states, whose spin configurations are noncoplanar. 
In the $3Q$ state, the constituent three spin helices are elliptical, as in the isotropic $2Q$ case in Sec.~\ref{sec:phasediagram2d}.
The calculated $R_S$, whose definition is the same as that in Fig.~\ref{fig14}(b)
(${\bf Q}_{\rm max}$ is ${\bf Q}_\eta$ for the largest $|{\bf S}_{{\bf Q}_{\eta}} |$), is plotted in Fig.~\ref{fig18}(b).
Note that $R_S=0$ in the isotropic $3Q$ state at $D=0$ where three sinusoidal spin density waves are superposed with equal weight. 
We find that $R_S$ increases while increasing $D$ and decreasing $\Delta$.

\subsubsection{Hedgehog lattices}\label{sec:hl3d}

\begin{figure}[!h]
  \centering
  \includegraphics[trim=0 0 0 0, clip,width=\columnwidth]{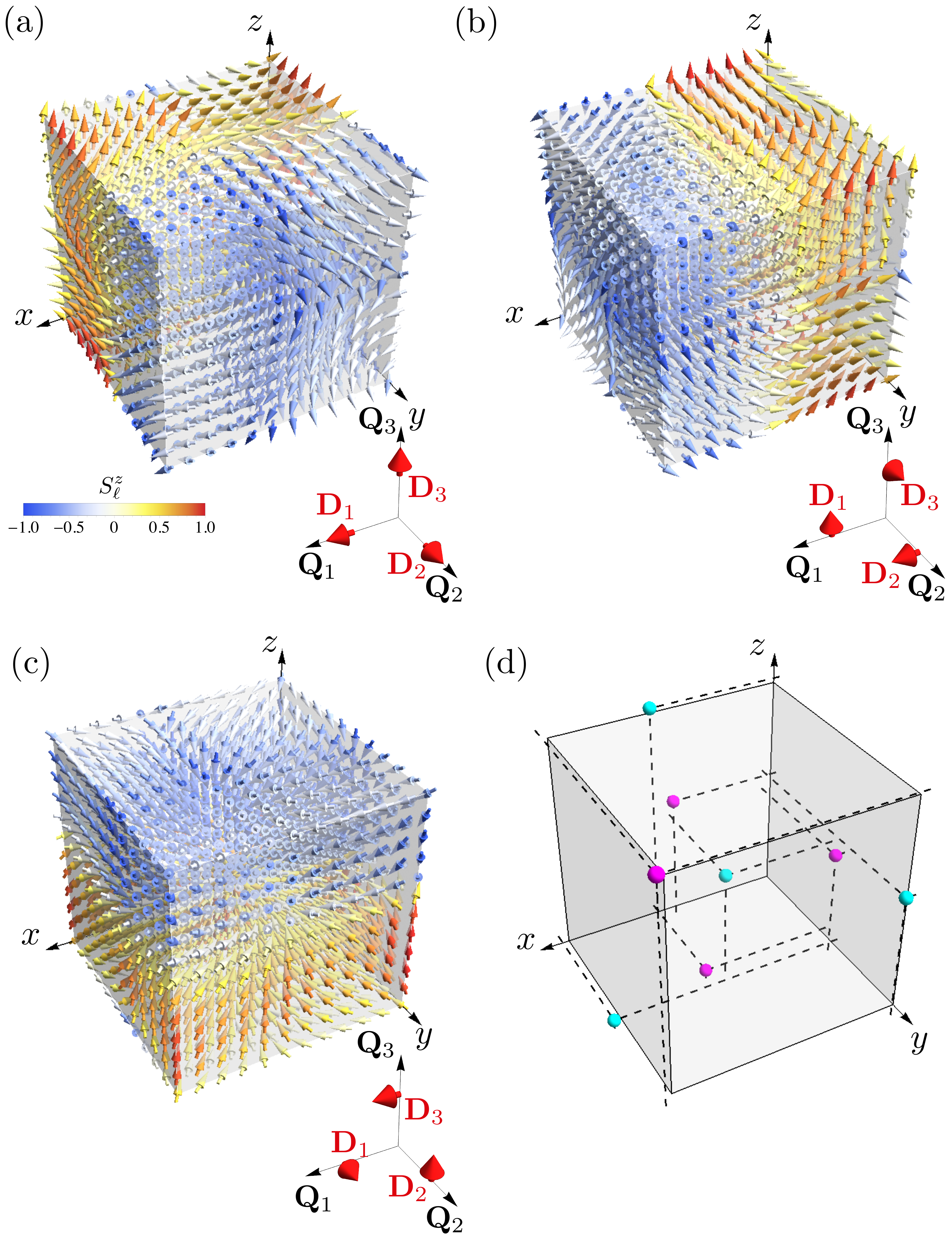}
  \caption{
  Isotropic $3Q$ spin states in the infinite-range model in three dimensions with $D/J=0.3$, $\Delta = 0.3$, and $\Lambda=12$:
  (a) proper-screw, (b) cycloid(I), and (c) cycloid(II) HLs. 
  The forms of $J^{\alpha\alpha}_{{\bf Q}_\eta}$ and ${\bf D}_{{\bf Q}_\eta}$ in each case are summarized in Table~\ref{tab1}.
  Insets show ${\bf D}_{{\bf Q}_\eta}$ as well as ${\bf Q}_\eta$ for each case.
  The color of arrows indicates the $z$ component of spin according to the color bar in (a).
  (d) Positions of the hedgehogs (magenta spheres) and the antihedgehogs (cyan spheres), which are common to all the HLs shown in (a)--(c). 
  The dashed lines are the guides for eyes.
  }
  \label{fig19}
\end{figure}
While the phase diagram is common to all the settings in Table~\ref{tab1}, the actual spin configuration in the $3Q$ state depends on the type of interactions. 
We present the variational results in Fig.~\ref{fig19}, focusing on the isotropic $3Q$ state at $D/J=0.3$ and $\Delta = 0.3$ ($R_S  \simeq 0.22$). 
Figure~\ref{fig19}(a) shows the stable spin configuration when taking $J^{\alpha\alpha}_{{\bf Q}_\eta}$ and ${\bf D}_{{\bf Q}_\eta}$ as Eqs.~\eqref{eq:JQeta3d1} and \eqref{eq:DQeta3d}, respectively. 
This is the proper-screw HL. 
On the other hand, Fig.~\ref{fig19}(b) shows the spin configuration obtained for Eqs.~\eqref{eq:JQeta3d2} and \eqref{eq:DQeta3d2}, which is the cycloid(I) HL. 
Likewise, Fig.~\ref{fig19}(c) displays the spin configuration for Eqs.~\eqref{eq:JQeta3d3} and \eqref{eq:DQeta3d3}, which is the cycloid(II) HL.
Figure~\ref{fig19}(d) shows the positions of the topological defects, i.e., hedgehogs and antihedgehogs,
which are identified as sources and sinks, respectively, of the emergent magnetic field defined by a solid angle formed by neighboring three spins~\cite{Okumura2020}.
The positions of the hedgehogs and antihedgehogs are common to the three HLs, since  
the spin configurations are mutually transformed
by $2\pi/3$ rotations about the $[111]$ axis in spin space as described in Sec.~\ref{sec:inf_3D}.

These HLs are stabilized by the anisotropy $\Delta$ in the symmetric interactions, as suggested in the phase diagram in Fig.~\ref{fig18}(a). 
Similar to the 2D case in Sec.~\ref{sec:vc2d}, this can be confirmed by calculating the spin components of ${\bf S}_{{\bf Q}_{\eta}}$. 
We find that all the HLs have the large amplitudes for 
$|S_{{\bf Q}_1}^y|=|S_{{\bf Q}_2}^z|=|S_{{\bf Q}_3}^x|$,
$|S_{{\bf Q}_1}^x|=|S_{{\bf Q}_2}^y|=|S_{{\bf Q}_3}^z|$,
 and $|S_{{\bf Q}_1}^z|=|S_{{\bf Q}_2}^x|=|S_{{\bf Q}_3}^y|$
for the proper-screw, cycloid(I), and cycloid(II), respectively,  
which gain the interaction energy in the presence of $\Delta$ in each case.

\subsubsection{Interaction range dependence}\label{sec:Int_range_dep3d}
\begin{figure}[htb]
  \centering
  \includegraphics[trim=0 0 0 0, clip,width=\columnwidth]{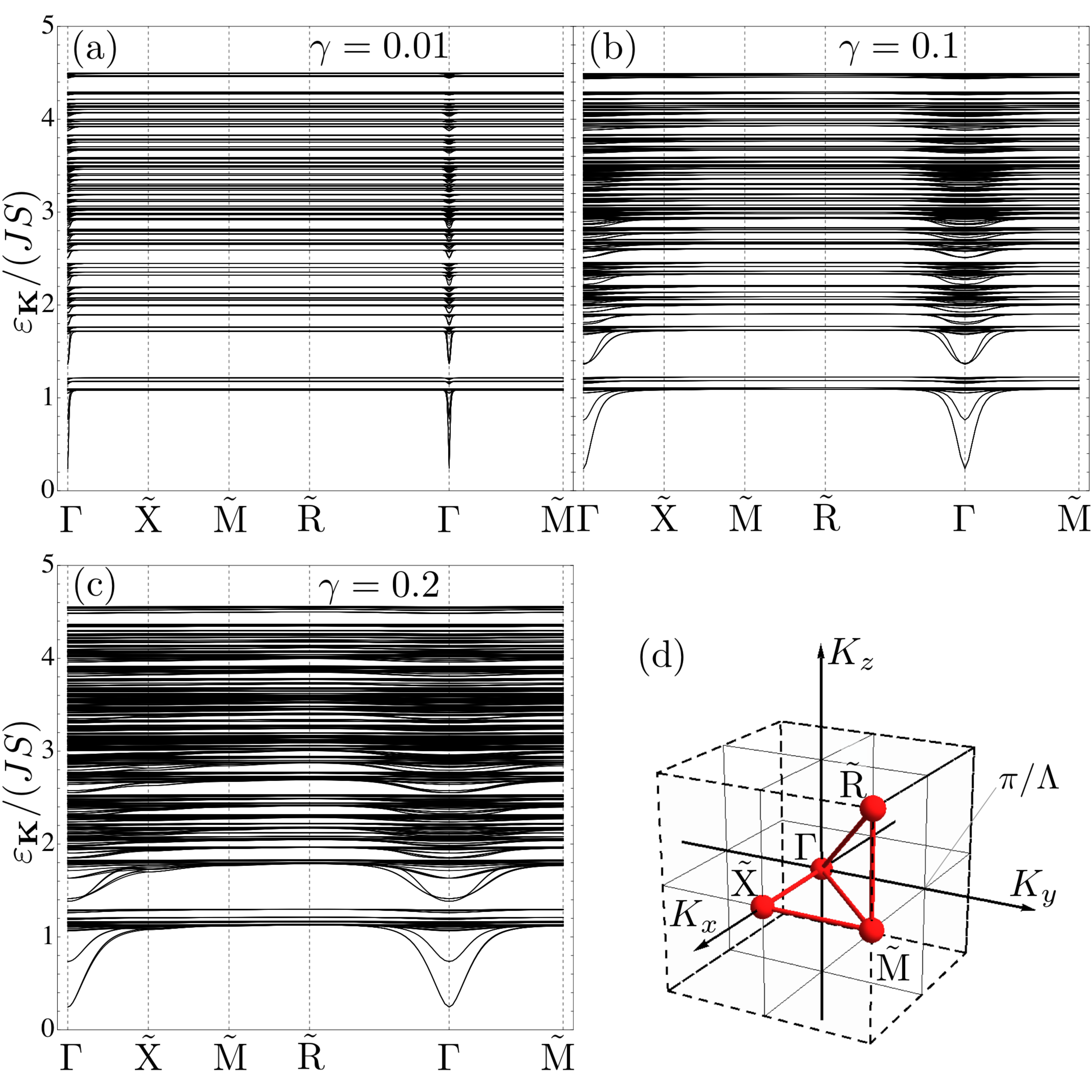}
  \caption{
    Spin excitation spectra in the isotropic $3Q$ spin states in three dimensions for 
    (a) $\gamma=0.01$, 
    (b) $\gamma=0.1$, and 
    (c) $\gamma=0.2$.
    We take $D/J=0.3$, $\Delta = 0.3$, and $\Lambda = 12$, as in Fig.~\ref{fig19}.
    (d) Symmetric lines in the folded Brillouin zone, used for the plots in (a--c).}
  \label{fig20}
\end{figure}
Let us discuss the finite-range model in three dimensions. 
As in the 2D case, we perform the variational calculations for $\gamma>0$ starting from the solutions for $\gamma=0$, and find stable but modified spin configurations.
Figure~\ref{fig20} shows the spin-wave dispersion $\varepsilon_{\bf K}$ in the folded Brillouin zone for three different values of $\gamma$ at $D/J=0.3$, and $\Delta = 0.3$ with $\Lambda = 12$; 
the results are common to all the HLs in Fig.~\ref{fig19}. 
Similar to the 1D and 2D cases, the excitation spectra are almost flat in most regions in momentum space for small $\gamma$, but they become more dispersive while increasing $\gamma$.

\subsubsection{Dynamical spin structure factor}
\label{sec:INS3d}
\begin{figure}[ht]
  \centering
  \includegraphics[trim=0 0 0 0, clip,width=\columnwidth]{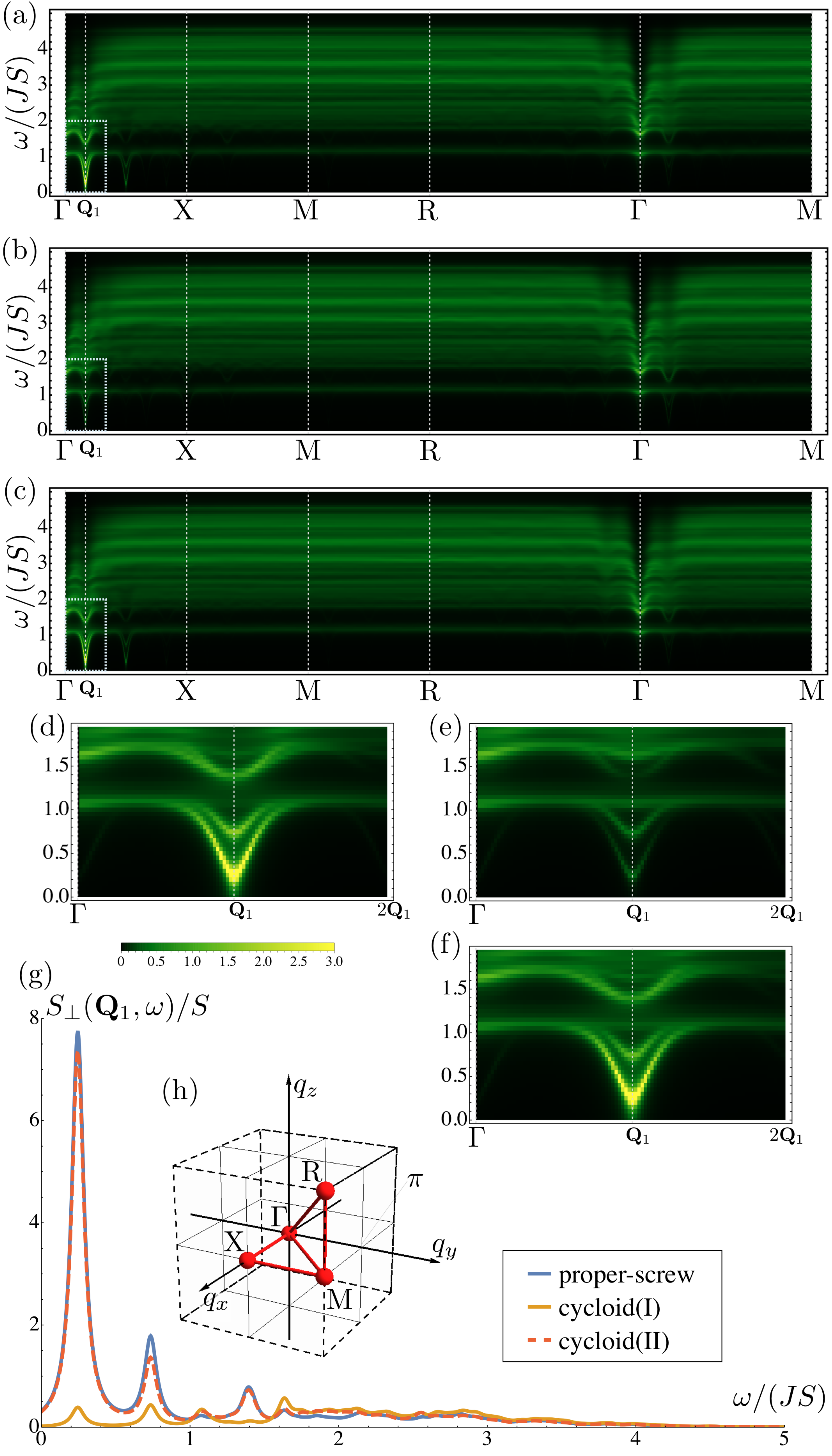}
  \caption{
Transverse component of the dynamical spin structure factor, $S_{\perp}({\bf q},\omega)$ in Eq.~\eqref{eq:Sqw_perp},
  for (a) proper-screw, (b) cycloid(I), and (c) cycloid(II) HLs, plotted along the symmetric lines in the first Brillouin zone shown in (h).
  (d), (e), and (f) show the enlarged views of the dotted areas in (a), (b), and (c), respectively.
  (g) $\omega$ dependence at ${\bf q}={\bf Q}_1$.
  The interaction range and the relaxation rate are taken as $\gamma=0.2$ and $\epsilon= 0.05$, respectively.
  The other parameters are $\Lambda = 12$, $D/J=0.3$, and $\Delta = 0.3$ as in Fig.~\ref{fig19}.
  }
  \label{fig21}
\end{figure}

Figure~\ref{fig21} shows the transverse component of the dynamical spin structure factor, $S_{\perp}({\bf q},\omega)$ in Eq.~\eqref{eq:Sqw_perp},
for the three types of HLs obtained at $D/J=0.3$, $\Delta = 0.3$, $\Lambda = 12$, and $\gamma=0.2$.
We take $\epsilon=0.05$ in Eq.~\eqref{eq:Sqw}. 
The calculations are done in a similar manner to the 2D case in Sec.~\ref{sec:INS2d}, by using the spin rotation as
\begin{align}
&{\bf S}=\left(
S^x, S^y, S^z
\right), ~ \text{proper-screw HL} \nonumber\\
&~~ \to
\begin{cases}
{\bf S}=\left(
S^y, S^z, S^x
\right), &\text{cycloid(I) HL}\\
{\bf S}=\left(
S^z, S^x, S^y
\right),
&\text{cycloid(II) HL}
\end{cases}.
\end{align}

The overall spectra look similar among the different HLs. 
In particular, as expected from the above calculation scheme, 
the spectra along the ${\rm R}$--$\Gamma$ line are common to all the three types of HLs.

On the other hand, similar to the 2D case, 
a stark difference among the three HLs 
is found in the vicinity of ${\bf q}={\bf Q}_1$:
The large intensities of the lowest-energy excitation mode are seen in the proper-screw and cycloid(II) HLs~[Figs.~\ref{fig21}(d) and \ref{fig21}(f)], 
whereas not in the cycloid(I) HL~[Fig.~\ref{fig21}(e)]. 
This is more clearly shown in the $\omega$ dependence at ${\bf q}={\bf Q}_1$ in Fig.~\ref{fig21}(g).
As in the 2D case, this large difference is consistently understood from $R_S^2$:
For the present parameter set, $R_S \simeq 0.22$ as shown in Sec.~\ref{sec:hl3d}, leading to $R_S^2 \approx 0.048$. 
Since the intensity for the cycloid(I) is particularly smaller than the other two, 
this difference could be useful to distinguish the cycloid(I) HL from the proper-screw and cycloid(II) HLs in experiments.

Although it is rather difficult to distinguish the proper-screw and cycloid(II) HLs solely form the spectra in Fig.~\ref{fig21}, 
it would be useful to separately measure the spin-flip and non-spin-flip cross sections in inelastic neutron scattering experiments. 
We expect a larger (non-)spin-flip component for the proper-screw [cycloid(II)] HL since it has a larger intensity in 
$S^{yy}({\bf Q}_1,\omega)$ [$S^{zz}({\bf Q}_1,\omega)$].

\section{Summary and discussion}\label{sec:summary_discussions}
We have investigated the spin excitation spectra for various types of helimagnetic states in the spin models with long-range exchange interactions. 
Starting from the model with infinite-range interactions, we have studied the models with long- but finite-range interactions including the symmetric diagonal ones with spin anisotropy and the antisymmetric off-diagonal ones of the Dzyaloshinskii-Moriya type. 
While changing the range of the interactions, we clarified the ground state and the spin excitation by the variational calculation and  the linear spin-wave theory, respectively. 
For the spin excitation, in addition to the spin-wave dispersion, we computed the transverse component of the dynamical spin structure factor, 
$S_\perp({\bf q},\omega)$, which is relevant to the inelastic neutron scattering experiments.

In the 1D case, we obtained the analytical solution for the spin excitation in the isotropic HS states with a spatially uniform twist angle, for both proper-screw and cycloid types. 
We showed that the spin-wave dispersion is completely flat in the infinite-range model except for $q=0$, $\pm Q$, and $\pm 2Q$, where $Q$ is the helical wave number, but it becomes dispersive for the finite-range case. 
Irrespective of the spatial range of interactions, there is a gapless excitation mode, which results in strong intensities at $q=\pm Q$ in the dynamical spin structure factor. 
Meanwhile, we also obtained the numerical results for the effect of the spin anisotropy in the symmetric diagonal interactions. 
We found that the anisotropy makes the twist angle of the stable 
spin texture inhomogeneous, 
and accordingly opens a gap in the spin-wave dispersion.
We also showed that, as long as the anisotropy is weak, the lowest-energy excitation mode can be regarded as a phase shift of the helix.
In addition, we found a discernible difference between the proper-screw and cycloid HS states 
in the high-energy spectra of $S_\perp({\bf q},\omega)$ at ${\bf q}=Q\hat{\bf x}$. 
We also found additional intensities in $S_\perp({\bf q},\omega)$ for the higher harmonics at $q=\pm 3Q$ in the presence of the spin anisotropy.

Extending the analyses to the 2D case, we have discussed the stability and excitations of $2Q$ VCs. 
By using the variational calculation, we found that the $2Q$ VCs are stabilized by the spin anisotropy. 
More specifically, while increasing the spin anisotropy, the $1Q$ HS state stabilized by the antisymmetric interactions turns into the $2Q$ VCs through the apparently continuous phase transition. 
There are two different $2Q$ VC phases: the anisotropic one with a superposition of two spin helices with different amplitudes and the isotropic one with equal amplitudes. 
The latter appears for larger anisotropy than the former.
While the phase diagram is common, the stable spin configurations of the VCs depend on the form of the interactions in the model, that is, 
the easy axes of the spin anisotropy and the directions of the Dzyaloshinskii-Moriya vectors. 
We obtained four types of VCs: two of them are superpositions of proper screws and the other two are of cycloids. 
By using the linear spin-wave theory, we showed that the spin-wave dispersion, which is common to the four VCs, becomes dispersive upon introducing the spatial decay of the interactions, similar to the 1D HS case. 
Meanwhile, we found discernible differences in $S_\perp({\bf q},\omega)$ among the four types of VCs at ${\bf q} \simeq {\bf Q}_1$ and ${\bf q} \simeq {\bf Q}_1 + {\bf Q}_2$. 
The finding could be useful to determine the type of VCs as well as the relevant effective spin model in inelastic neutron scattering experiments.

In the 3D case, we have examined the stable spin configurations and the spin excitation spectra for three types of $3Q$ HLs: one proper-screw type and two cycloid types. 
In the variational phase diagram, which is common to the three HLs, we found that the HLs are stabilized in the presence of the spin anisotropy, similar to the $2Q$ VCs in the 2D case. 
In this 3D case, however, while increasing the spin anisotropy, the $1Q$ HS state first turns into the anisotropic $2Q$ VC, and then into the isotropic $3Q$ HL; 
the phase transition between the $2Q$ VC and the $1Q$ HS state looks continuous,
while that between the $3Q$ HL and the $2Q$ VC is discontinuous. 
With regard to the spin excitation spectra, we found qualitatively similar behaviors to the $2Q$ VC cases. 
Thus, in this case also, the differences in $S_\perp({\bf q},\omega)$ would be useful to distinguish the type of HLs and to identify the relevant interactions in experiments.

Finally, we discuss candidate materials to which our results are potentially relevant.
As discussed in Sec.~\ref{sec:anisotropy1d}, the 1D HS states in CuB$_2$O$_4$ and TbMnO$_3$ could be accounted for by our 
1D model with the spin anisotropy, which predicts higher harmonics at $q=\pm 3Q$ in the dynamical spin structure factor 
as observed in the experiments~\cite{Roessli2001,Kajimoto2004}.
For the 2D (3D) models, 
the magnetic metals with the crystallographic point groups $D_4$, $C_{4v}$, and $D_{2d}$ ($T$ and $C_3$) can be candidate materials, as shown in Table~\ref{tab1}:
for example, a Mn-Pt-Sn inverse Heusler compound ($D_{2d}$) where an antiskyrmion crystal has been found~\cite{Nayak2017},
and MnSi$_{1-x}$Ge$_{x}$ of $B20$ structure ($T$) where the magnetic HLs have been found~\cite{Tanigaki2015,Kanazawa2016,Kanazawa2017,Fujishiro2019,Fujishiro2020,Kanazawa2020}.
To the best of our knowledge, 
inelastic neutron scattering experiments have not been performed systematically for the multiple-$Q$ spin states of VC and HL thus far.
We hope that our results stimulate such experiments and the detailed comparison between theory and experiment provides a hint for understanding of the microscopic mechanism of the multiple-$Q$ states.

In addition to the above substances, recently, multiple-$Q$ spin states in centrosymmetric systems, for which the antisymmetric interactions of the Dzyaloshinskii-Moriya type are inactive, have been attracting considerable attentions,
for example, GdRu$_2$Si$_2$ ($D_{4h}$)~\cite{Khanh2020,Yasui2020}, 
Gd$_2$PdSi$_3$ ($D_{6h}$)~\cite{Kurumaji2019,Sampathkumaran2019,Hirschberger2020,Hirschberger2020b,Kumar2020,Spachmann2021}, 
Gd$_3$Ru$_4$Al$_{12}$ ($D_{6h}$)~\cite{Hirschberger2019,Hirschberger2021},
and SrFeO$_3$ ($O_h$)~\cite{Ishiwata2011,Rogge2019,Ishiwata2020,Onose2020}.
In our model, however, when the antisymmetric interactions are absent, the ground states are almost always given by superpositions of sinusoidal spin density waves, 
inconsistent with the experimental observations.
Thus, for these multiple-$Q$ spin states, further extensions of the model are necessary, e.g., additional biquadratic interactions~\cite{Okumura2020,Shimizu2021a,Shimizu2021b}.
This interesting issue is left for future research.

\begin{acknowledgments}
The authors thank 
Y. Fujishiro,
N. Kanazawa, and 
T. Nakajima
for fruitful discussions.
This work was partially supported by Japan Society for the Promotion of Science (JSPS) KAKENHI Grant No. JP18K03447, JP19H01834, and JP19H05825, JST CREST Grant No. JPMJCR18T2,
and
JST PRESTO (JPMJPR20L8).
\end{acknowledgments}

\appendix
\section{Symmetry argument for $S_\perp({\bf q},\omega)$ along the ${\rm X}$--${\rm M}$ line in 2D VCs}\label{sec:appA}
In this Appendix, we explain why $S_\perp({\bf q},\omega)$ along the ${\rm X}$--${\rm M}$ line are common to the two types of VCs for both proper-screw and cycloidal cases as shown in Fig.~\ref{fig15}, from a symmetry argument.
Let ${\bf q}_0$ be any wavenumber on the ${\rm X}$--${\rm M}$ line.
$S_\perp({\bf q},\omega)$ at ${\bf q}={\bf q}_0$ for the type (I) VCs are computed by using Eq.~\eqref{eq:Sqw} as
\begin{align}
S_{\perp}^{{\rm I}} ({\bf q}_0 , \omega) = S^{zz}({\bf q}_0, \omega) + S^{\mu_2\mu_2}({\bf q}_0, \omega),
\end{align}
where 
the spin component in the second term is taken along $\hat{\bm \mu}=(-\hat{q}^y_0, \hat{q}^x_0,0)$; here,
$\hat{\bf q}_0 = (\hat{q}^x_0,\hat{q}^y_0)= {\bf q}_0/|{\bf q}_0|$.
On the other hand, those for the type (II) VCs, $S^{\rm II}_\perp({\bf q}_0,\omega)$, are computed by replacing $\hat{\bm \mu}$ 
by $\hat{\bm \mu}'=(\hat{q}^y_0, \hat{q}^x_0,0)$ because
the type (I) and (II) VCs are connected each other by $\pi$ rotation about $[100]$ axis in spin space.
Denoting ${\bf q}'_0= (-q^x_0,q^y_0) \perp (\hat{\mu}'{}^x , \hat{\mu}'{}^y)$, 
which is connected to ${\bf q}_0$ by a reciprocal vector [${\bf q}_0' = {\bf q}_0 -(2\pi,0)$], 
we obtain 
\begin{align}
S^{\rm II}_\perp ({\bf q}_0,\omega) = 
S^{\rm I}_\perp ({\bf q}_0', \omega). \label{eq:SS1}
\end{align}
In addition, the following relation holds:
\begin{align}
S^{\rm I}_\perp 
({\bf q}_0,\omega) = 
S^{\rm I}_\perp 
({\bf q}_0', \omega),\label{eq:SS2}
\end{align}
because the Hamiltonian and the ground-state spin configurations are invariant under
$\pi$ rotation about the $[010]$ axis in both coordinate and spin spaces for the proper-screw(I) VC and under a combined operation of $\pi$ rotation about the $[010]$ axis in coordinate space and
$\pi$ rotation about the $[100]$ axis in spin space for the cycloid(I) VC.
Finally, from Eqs.~\eqref{eq:SS1} and \eqref{eq:SS2}, we find
\begin{align}
S^{\rm I}_\perp ({\bf q}_0,\omega) = 
S^{\rm II}_\perp ({\bf q}_0, \omega),
\end{align}
where ${\bf q}_0$ is in the ${\rm X}$--${\rm M}$ line.



\break
\bibliography{draft} 

\end{document}